\documentclass[twocolumn]{jpsj2} %% two-column layout
\def\eg{e.g.}

\def\etal{{\it et\ al.}}

\newcommand{\lsim} 
 {\ \raise.35ex\hbox{$<$}\kern-0.75em\lower.5ex\hbox{$\sim$}\ }
\newcommand{\gsim}
 {\ \raise.35ex\hbox{$>$}\kern-0.75em\lower.5ex\hbox{$\sim$}\ }
\def\journal #1#2#3#4{#1 {\bf #2} (#4) #3}
\def\PR{Phys.\ Rev.}

\def\PRB{Phys.\ Rev.\ B}
\def\PRL{Phys.\ Rev.\ Lett.}

\def\JPSJ{J.\ Phys.\ Soc.\ Jpn.}

\def\RMP{Rev.\ Mod.\ Phys.}
\def\PTP{Prog.\ Theor.\ Phys.}

\def\ZPB{Z.\ Phys.\ B}

\def\ACB{Acta.\ Cryst.\ B}
\def\COMP{Comments~Cond.~Mat.~Phys.}
\title{
Superconductivity and a Mott Transition 
in a Hubbard Model on an Anisotropic Triangular Lattice
}% Force line breaks with \\

\author{Tsutomu \textsc{Watanabe}$^{1,2}$\thanks{E-mail: 
h042203d@mbox.nagoya-u.ac.jp}, Hisatoshi \textsc{Yokoyama}$^{3}$, 
Yukio \textsc{Tanaka}$^{1,2}$ and Jun-ichiro \textsc{Inoue}$^{1}$ 
}

\inst{
$^{1}$Department of Applied Physics, Nagoya University,  Nagoya 464-8603 \\
$^{2}$CREST Japan Science and Technology Corporation (JST) \\
$^{3}$Department of Physics, Tohoku University, Sendai 980-8578 \\
}

\abst{
A half-filled-band Hubbard model on an anisotropic triangular 
lattice ($t$ in two bond directions and $t'$ in the other) 
is studied using an optimization variational Monte Carlo method, 
to consider a Mott transition and superconductivity 
arising in $\kappa$-$({\rm BEDT}$-${\rm TTF})_2$X. 
Adopting wave functions with doublon-holon binding factors, 
we reveal that a first-order Mott (conductor-to-nonmagnetic 
insulator) transition takes place at $U=U_{\rm c}$, approximately 
of the bandwidth, for a wide $t'/t$ range. 
This transition is not directly connected to magnetism. 
Robust $d$-wave superconductivity appears in a restricted 
parameter range: immediately below $U_{\rm c}$ and the moderate strength 
of frustration ($0.4\lsim t'/t\lsim 0.7$), where short-range 
antiferromagnetic correlation sufficiently develops but does 
not come to a long-range order. 
The relevance to experiments is also discussed. 
}

\kword{Hubbard model, anisotropic triangular lattice, 
variational Monte Carlo method, superconductivity, 
$d_{x^2-y^2}$ wave, antiferromagnetic correlation,
Mott transition, $\kappa$-$({\rm BEDT}$-${\rm TTF})_2$X}

\begin{document}
\maketitle

%%%%%%%%%%%%%%%%%%%%%%%%%%%%%%%%%%%%%%%%%%%%%%%%%%%%%%%%%%%%%%%%%%%%%%%%%
\section{Introduction\label{sec:intro}}
%%%%%%%%%%%%%%%%%%%%%%%%%%%%%%%%%%%%%%%%%%%%%%%%%%%%%%%%%%%%%%%%%%%%%%%%%
Organic layered compounds made of BEDT-TTF (ET) molecules present 
various characteristics of correlated electron systems. 
Among them, $\kappa$-type compounds \cite{ETKa,ETMc,bandMc}, 
$\kappa$-${\rm (ET)_2}$X (X denotes an anion), have intriguing properties 
as follows: 
(1) They have good two-dimensionality in conductivity with a frustrated 
lattice structure. 
(2) Most of them show unconventional (probably 
$d_{x^2-y^2}$-wave) superconductivity (SC) as high as 10 K, with 
pseudo-gap behavior for $T>T_{\rm c}$. 
These two points are closely akin to high-$T_{\rm c}$ cuprates. 
(3) A series of $\kappa$-(ET) salts give rise to 
superconductor-to-antiferromagnetic (AF) insulator transitions, 
through the chemical substitution of X, or applied pressure. 
(4) It was recently found that a compound [X=Cu$_2$(CN)$_3$] preserves 
an insulating state without magnetic order down to a quite low 
temperature (23 mK). \cite{Shimizu}
\par 

About a decade ago, Kino and Fukuyama \cite{Fukuyama} studied the 
electronic structure of ET salts, applying a Hartree-Fock approximation 
to four-band Hubbard models, and found that if the dimerization 
between ET molecules is sufficiently strong, the conducting plane of 
$\kappa$-type compounds is described well by a single-band Hubbard 
model on an anisotropic triangular lattice (the hopping integral is 
$t$ in two bond directions and $t'$ in the other) with 
the electron density of half filling. 
The Hartree-Fock approximation can describe a metal-to-AF insulator 
transition. 
Afterward, this model has been studied with various theories. 
Fluctuation exchange approximation (FLEX) \cite{Kino,Kondo} and 
a quantum Monte Carlo method \cite{Kuroki1} have yielded a result 
that SC with a $d_{x^2-y^2}$-wave symmetry is realized due to 
AF spin correlation (or fluctuation), as in the case of cuprates. 
Actually, experimental studies using NMR \cite{NMRMa,NMRSo,NMRKa}, 
thermal conductivity \cite{Izawa}, and tunneling spectroscopy measurements 
\cite{Arai,Ichimura,Tanuma} have found unconventional SC with 
nodes of the gap in $\kappa$-${\rm (ET)_2}$X. 
\par

However, it is difficult to deal with essential aspects of a 
strong correlation, particularly Mott transitions \cite{Mott} 
by Hartree-Fock or FLEX approximation.  
In contrast, dynamical cluster approximation (DCA) \cite{DCA} 
and path integral renormalization group calculation (PIRG) 
\cite{Morita} have shown first-order metal-to-nonmagnetic-insulator 
transitions for intermediately and highly frustrated cases. 
However, both DCA and PIRG have not reached a definite conclusion 
regarding SC in the vicinity of Mott transitions. 
Thus, it is important to clarify the interrelationship between 
SC and a Mott transition. 
\par

The purpose of this paper is to understand simultaneously SC and 
a Mott transition arising in the Hubbard model on an anisotropic 
triangular lattice at half filling, with $\kappa$-ET salts in mind. 
To this end, we use a variational Monte Carlo (VMC) method, 
which is very useful in studying the physics of intermediate 
correlation strength as a continuous function of $U/t$. \cite{YTOT} 
In the variational wave functions, we introduce a factor 
controlling the binding between a doublon and a holon, which 
is necessary to appropriately treat the Hubbard model including 
Mott transitions. \cite{YokoPTP}
We carefully check the system-size dependence with highly 
accurate VMC data; this check is indispensable for the study 
of phase transitions. 
\par

It is found that a first-order Mott transition takes place 
at $U=U_{\rm c}$ approximately of the bandwidth for a wide $t'/t$ 
range. 
The $d$-wave superconducting (SC) correlation function indicates 
that robust SC is realized immediately below $U_{\rm c}$ for a moderate 
frustration ($0.4\lsim t'/t\lsim 0.7$). 
For weakly frustrated cases ($t'/t\lsim 0.4$), the SC is 
defeated by an AF state with a long-range order; for highly 
frustrated cases ($t'/t\gsim 0.8)$, the SC correlation becomes weak 
for arbitrary values of $U/t$, together with a decrease in the degree 
of AF correlation. 
Thus, the origin of the SC can be attributed to the AF correlation. 
\par

The organization of this paper is as follows: 
In \S\ref{sec:model}, the model and method used are introduced. 
Sections \ref{sec:comparison}-\ref{sec:pair} are assigned to the 
description of our results and comparisons with other theories. 
Section \ref{sec:comparison} treats the behavior of condensation 
energies of the $d$-wave and AF states. 
In \S\ref{sec:mott}, we demonstrate that a Mott transition actually 
takes place in the present approach by studying various quantities, 
and reveal the properties of this Mott transition. 
In \S\ref{sec:pair}, we describe robust $d$-wave SC and its relevance 
to the AF spin correlation. 
\S\ref{sec:discussions} presents further discussions; first 
we compare our results with those of related experiments on $\kappa$-ET 
salts, followed by an analysis of the Mott insulating state 
in light of strong-coupling theories. 
In \S\ref{sec:summary}, we summarize our results. 
In Appendix A, we point out the limitation of Gutzwiller-type 
approximations in dealing with a Mott transition. 
In Appendix B, we comment on the related previous variational 
studies in the following context. 
Although a couple of studies in the early 90's\cite{YSMott,Millis} 
concluded that a Mott transition never arises in such 
wave functions as treated in this paper, this conclusion 
has already been proven incorrect.\cite{YokoPTP} 
However, a recent VMC study on the present subject \cite{Liu} 
has still followed the above wrong conclusion and made some 
misinterpretations. 
\par 

Part of this study has been reported in a proceedings. 
\cite{WataISS}

%%%%%%%%%%%%%%%%%%%%%%%%%%%%%%%%%%%%%%%%%%%%%%%%%%%%%%%%%%%%%%%%%%%%%%%%%
\section{Model and Method\label{sec:model}}
%%%%%%%%%%%%%%%%%%%%%%%%%%%%%%%%%%%%%%%%%%%%%%%%%%%%%%%%%%%%%%%%%%%%%%%%%

%************************************************************************
% Model
%************************************************************************
\begin{figure}
\begin{center}
\includegraphics[width=5.5cm,height=3.5cm]{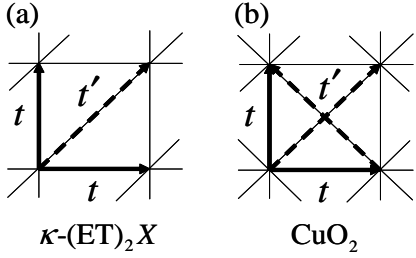}
\end{center}
\caption{
Lattice structure and hopping integrals of models (a) used in this 
study for $\kappa$-ET salts, and (b) often used for high-$T_{\rm c}$ 
cuprates. 
The latter is shown for comparison. 
}
\label{fig:model}
\end{figure}
%************************************************************************

According to the extended H\"uckel calculations based on the 
structure of $\kappa$-ET salts \cite{bandKo,bandMc}, 
their complicated conducting planes are described with an 
anisotropic triangular lattice, whose unit cell is composed of 
a dimer of ET molecules. 
In each dimer (or unit cell), the strong hybridization between 
two ET molecules leads to a large split of the levels of 
bonding and antibonding orbitals. 
The antibonding orbital accommodates one electron, and composes 
a conduction band of half filling. 
Because the Coulomb interaction is relatively strong due to the 
slight overlap between $\pi$ orbitals, a Hubbard model 
has been considered as a realistic model.\cite{Tamura,Fukuyama}
\par

We hence consider a single-band Hubbard model on an anisotropic 
triangular lattice [see Fig.\ref{fig:model}(a)] given as 
%----------------------------------------------------------------------
\begin{eqnarray}
{\cal H}={\cal H}_{\rm kin}+{\cal H}_{\rm int}=
   \sum_{{\bf k}\sigma} \varepsilon_{\bf k} 
   c_{{\bf k}\sigma}^\dag c_{{\bf k}\sigma}+U\sum 
   \limits_i n_{i\uparrow}n_{i\downarrow}, 
\label{eq:model} 
\end{eqnarray}
%----------------------------------------------------------------------
\begin{equation} 
\varepsilon_{\bf k}=-2t(\cos k_x+\cos k_y)-2t'\cos (k_x+k_y),
\label{eq:dispersion}
\end{equation}
%----------------------------------------------------------------------
with $U,t,t'>0$.\cite{comment1} 
Throughout this paper, we fix the electron density at half filling
($n=N_{\rm e}/N_{\rm s}=1$; $N_{\rm e}$: electron number, 
$N_{\rm s}$: site number). 
\par

To this model, we apply an optimization VMC method \cite{Umrigar} 
that can correctly treat the local correlation in the whole parameter 
space spanned by $U/t$ and $t'/t$. 
As a variational wave function, we use a Jastrow type: 
$\Psi={\cal P}\Phi$, 
in which $\Phi$ is a one-body (Hartree-Fock) part expressed as 
a Slater determinant, and $\cal P$ is a many-body correlation factor. 
Concerning $\cal P$, the onsite (Gutzwiller) factor\cite{Gutzwiller} 
\begin{equation} 
{\cal P}_{\rm G}=\prod_i\left[1-(1-g) n_{i\uparrow}
n_{i\downarrow}\right], 
\end{equation} 
was found effective for $t$-$J$-type models. \cite{YOtJ,Paramekanti} 
For the Hubbard model, however, it has been known\cite{YS} that 
the wave functions ${\cal P}_{\rm G}\Phi(=\Psi_{\rm G})$ yield several 
unfavorable results, especially for large values of $U/t$: 
(1) The momentum distribution function $n({\bf k})$ is an increasing 
function of $|{\bf k}|$. 
(2) $2k_{\rm F}$ anomalies in the spin [charge density] structure 
factor $S({\bf q})$ [$N({\bf q})$] cannot be properly represented. 
(3) A Mott transition cannot be described, in addition to a sizably 
high variational energy. 
\par

%*****************************************************************************
% doublon-holon binding
%*****************************************************************************
\begin{figure}
\begin{center}
\includegraphics[width=7.0cm,height=2.5cm]{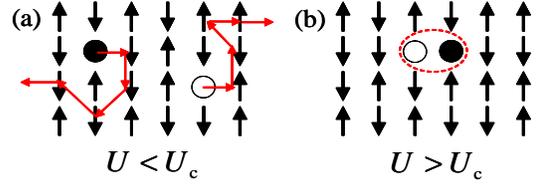} 
\end{center}
\vspace{-0.5cm}
\caption{(Color online) 
Schematic explanation of a Mott transition due to 
binding of a doublon to a holon described by $\Psi_Q^d$. 
The solid (open) circles denote doublons (holons). 
(a) In a conductive regime, a doublon and a holon (charge carriers) can 
move independently. 
(b) In an insulating regime, a doublon and a holon are confined within 
adjacent sites; thereby, charge cannot flow. 
The results of Mott transitions are treated in \S\ref{sec:mott}.
}
\label{fig:d-h2}
\end{figure}
%*****************************************************************************

To overcome these flaws, one has to introduce an intersite correlation 
factor. \cite{YSMott,Sorella}
In particular, at half filling, the doublon-holon binding factor 
${\cal P}_Q$\cite{Kaplan,YSMott,YokoPTP}, in addition to 
${\cal P}_{\rm G}$, is an indispensable element for describing 
a Mott transition. 
Generally, ${\cal P}_Q$ should be written as a functional of 
$\mu({\bf r})$, which is a function of the relative position 
of a doubly occupied site (doublon) from an empty site (holon), 
${\bf r}$. \cite{YSMott}
Here, however, we integrate only the two major elements into ${\cal P}_Q$, 
namely, for nearest neighbors, $\mu$ [${\bf r}=(\pm x,0)$ and $(0,\pm y)$] 
and for second (diagonal) neighbors, $\mu'$ [${\bf r}=(x,y)$ and $(-x,-y)$]. 
%---------------------------------------------------------------------
\begin{eqnarray}
{\cal P}_Q=
\prod_i{\bigl(1-\mu Q_i^{\tau} \bigr)
\bigl(1-\mu'Q_i^{\tau'}\bigr)} 
\label{eq:PQ}
\end{eqnarray}
%----------------------------------------------------------------------
%----------------------------------------------------------------------
\begin{eqnarray}
Q_i^{\tau(\tau')}=\prod_{\tau(\tau')} 
\left[d_i(1-e_{i+\tau(\tau')})+e_i(1-d_{i+\tau(\tau')})\right] 
\label{eq:Qi}
\end{eqnarray} 
%----------------------------------------------------------------------
Here, $d_i=n_{i\uparrow}n_{i\downarrow}$, 
$e_i=(1-n_{i\uparrow})(1-n_{i\downarrow})$, and $\tau$ ($\tau'$) 
runs over all adjacent sites in the bond directions of $t$ ($t'$). 
Our trial wave function thus becomes 
\begin{equation}
\Psi_Q={\cal P}_Q{\cal P}_{\rm G}\Phi. 
\label{eq:wf}
\end{equation}
The factor ${\cal P}_Q$ is naturally derived by taking account of 
virtual states in the second-order processes in the strong-coupling 
expansion,\cite{commentPQ,Harris} and controls the binding strength 
between a doublon and a holon with the parameters $\mu$ and $\mu'$ 
($\mu,\ \mu'\le 1$). 
For $\mu (\mu')\sim 0$, doublons and holons, which are charged, move 
around almost freely, whereas for $\mu (\mu') \rightarrow 1$, a doublon 
and a holon are tightly bound within adjacent sites in the bond 
directions of $t$ ($t'$) (See Fig.~\ref{fig:d-h2}). 
Thus, current cannot flow for $\mu\rightarrow 1$. 
It has been argued in detail \cite{YokoPTP} that this factor ${\cal P}_Q$ 
can describe a Mott transition and its existence has been actually 
confirmed for a regular square lattice. 
Afterward, Mott transitions have been found using ${\cal P}_Q$ 
for various systems (refer to \S\ref{sec:exp} in Appendix B). 
\cite{YTOT,YTOnew,kagome,CB,twoband}
\par

Regarding $\Phi$, we treat three cases in this paper, a $d$-wave BCS state, 
an AF state and a Fermi sea. 
First, we introduce a singlet wave function of the form of the BCS function 
with a $d_{x^2-y^2}$-wave gap 
%----------------------------------------------------------------------
\begin{equation}
\Phi_d =\left(\sum_{\bf k}\varphi_{\bf k}
c_{{\bf k}\uparrow}^\dagger c_{{\bf -k}\downarrow}^\dagger
\right)^\frac{N_{\rm e}}{2}|0\rangle, 
\label{eq:singlet}
\end{equation}
%----------------------------------------------------------------------
%----------------------------------------------------------------------
\begin{equation}
\varphi_{\bf k}=\frac{\Delta_{\bf k}}
{\tilde\varepsilon_{\bf k}-\zeta+
\sqrt{(\tilde\varepsilon_{\bf k}-\zeta)^2+
\Delta_{\bf k}^2}}. 
\label{eq:BCSDelta}
\end{equation}
%----------------------------------------------------------------------
with 
\begin{equation}
\tilde\varepsilon_{\bf k}=-2t(\cos k_x+\cos k_y)-2\tilde t'\cos (k_x+k_y), 
\end{equation}
and $\Delta_{\bf k}=\Delta_d(\cos k_x-\cos k_y)$.  
Here, $\zeta$, $\Delta_d$, and $\tilde t'$ are variational parameters 
to be optimized and $\zeta$ corresponds to the chemical potential. 
Although the parameter $\Delta_d$ indicates the magnitude of 
the $d$-wave gap, it does not necessarily mean the stability of 
SC; in the insulating phase, $\Delta_d$ is 
related to the spin gap. 
Furthermore, we take account of a renormalization effect of the 
quasi-Fermi surface due to the electron correlation. 
The main part of this effect is introduced into $\Phi_d$ by optimizing 
$\tilde t'/t$, independently of $t'/t$ given in the Hamiltonian 
eq.~(\ref{eq:model}). \cite{Himeda-t'}
As we will see later, $\tilde t'/t$ notably deviates 
from the original $t'/t$, particularly in the insulating phase, 
and the renormalization affects the stability of the $d$-wave singlet 
state. 
\par

Next, as a competing state for weakly frustrated (small-$t'/t$) cases, 
we consider the commensurate AF long-range ordered state 
%----------------------------------------------------------------------
\begin{equation}
\Phi_{\rm AF} =\prod\limits_{{\bf k},\sigma} 
{(\tilde u_{\bf k} c_{{\bf k}\sigma}^\dag + 
{\mathop{\rm sgn}}(\sigma)\tilde v_{\bf k} 
c_{{\bf k}+{\bf K}\sigma}^\dag)} 
\left|0\right\rangle, 
\end{equation}
%----------------------------------------------------------------------
with 
%----------------------------------------------------------------------
\begin{equation}
\tilde u_{\bf k} (\tilde v_{\bf k}) = 
\sqrt{\frac{1}{2}\left({1-(+)\frac{{\varepsilon_{\bf k}}}
{{\sqrt{\varepsilon_{\bf k}^2 + \Delta_{\rm AF}^2 }}}}\right)}. 
\end{equation}
%----------------------------------------------------------------------
Here, ${\bf K}=(\pi,\pi)$, and the wave-number space ($\bf k$) is 
restricted to the folded AF Brillouin zone. 
$\Delta_{\rm AF}$ is a variational parameter indicating the magnitude 
of the AF gap. 
At half filling, 
$\Psi_{\rm AF}(={\cal P}_Q{\cal P}_{\rm G}\Phi_{\rm AF})$ with finite 
$\Delta_{\rm AF}$ is always insulating, as in the mean-field theory. 
\par

Lastly, as a reference state, we use a projected Fermi sea: 
$\Psi_{\rm F}(={\cal P}_Q{\cal P}_{\rm G}\Phi_{\rm F})$ with 
$\Phi_{\rm F}=\prod_{|{\bf k}|<k_{\rm F}\sigma} 
c^\dag_{{\bf k}\sigma}|0\rangle$. 
Although $\Psi_{\rm F}$ also brings about a Mott transition at 
a somewhat larger $U$ than the bandwidth, \cite{YokoPTP,YTOnew} 
we will not treat the insulating regime of $\Psi_{\rm F}$ 
in this paper, and thus call it the `normal' or `metallic' state. 
\par

In the remainder of this section, we briefly mention the details of 
VMC calculations. 
Because the trial functions we treat have at most six parameters, 
we employ a simple linear optimization of each parameter with 
the other parameters fixed. 
In a round of iteration, every parameter is once optimized in one
dimension. 
In almost all the cases of optimization in this study, the parameters 
converge within five rounds, after which we continue the optimization 
for more than ten rounds. 
The optimized values of the parameters and energy are determined 
by averaging the results of the ten rounds after convergence. 
Because the optimization is carried out with $2$-$5 \times 10^5$ samples, 
our data are substantially averages of several million samples. 
Thus, we can greatly suppress the statistical errors in energy 
down to the order of $10^{-4}t$. 
Such precision is necessary to correctly analyze the physics in the 
vicinity of a phase transition. 
Physical quantities are calculated with the optimized parameters 
thus obtained with $2.5\times 10^5$ samples. 
\par

The systems used are of $N_{\rm s} = L \times L$ sites with 
periodic-antiperiodic boundary conditions. 
A closed shell condition is always imposed. 
We mainly study the systems for $L=10$ and $12$, and check those for 
$L=14$ and 16 when we would like to know the system-size dependence 
more definitely. 
As we will see later, for $t'\ne 0$, in contrast to for $t'=0$, the 
system-size dependence is not monotonic; in particular, the 
SC correlation function is very sensitive to the configuration of 
occupied $k$ points. 
Thus, we need to deduce the properties for $L\rightarrow\infty$ from 
the tendency of finite-sized systems in varying $L$. 
\par

%%%%%%%%%%%%%%%%%%%%%%%%%%%%%%%%%%%%%%%%%%%%%%%%%%%%%%%%%%%%%%%%%%%%%%%%%%
\section{Stability of $d$-wave and AF states\label{sec:comparison}}
%%%%%%%%%%%%%%%%%%%%%%%%%%%%%%%%%%%%%%%%%%%%%%%%%%%%%%%%%%%%%%%%%%%%%%%%%%

In this and the following two sections, we describe results of 
VMC calculations, and compare with other theoretical ones. 
In this section, we first consider the competition between $d$-wave and 
AF states, and then mention the critical behavior found for $\Psi_Q^d$. 
\par

%**********************************************************************
% Condensation energy
%**********************************************************************
\begin{figure*}[!t]
\begin{center}
\includegraphics[width=15cm,height=10.5cm]{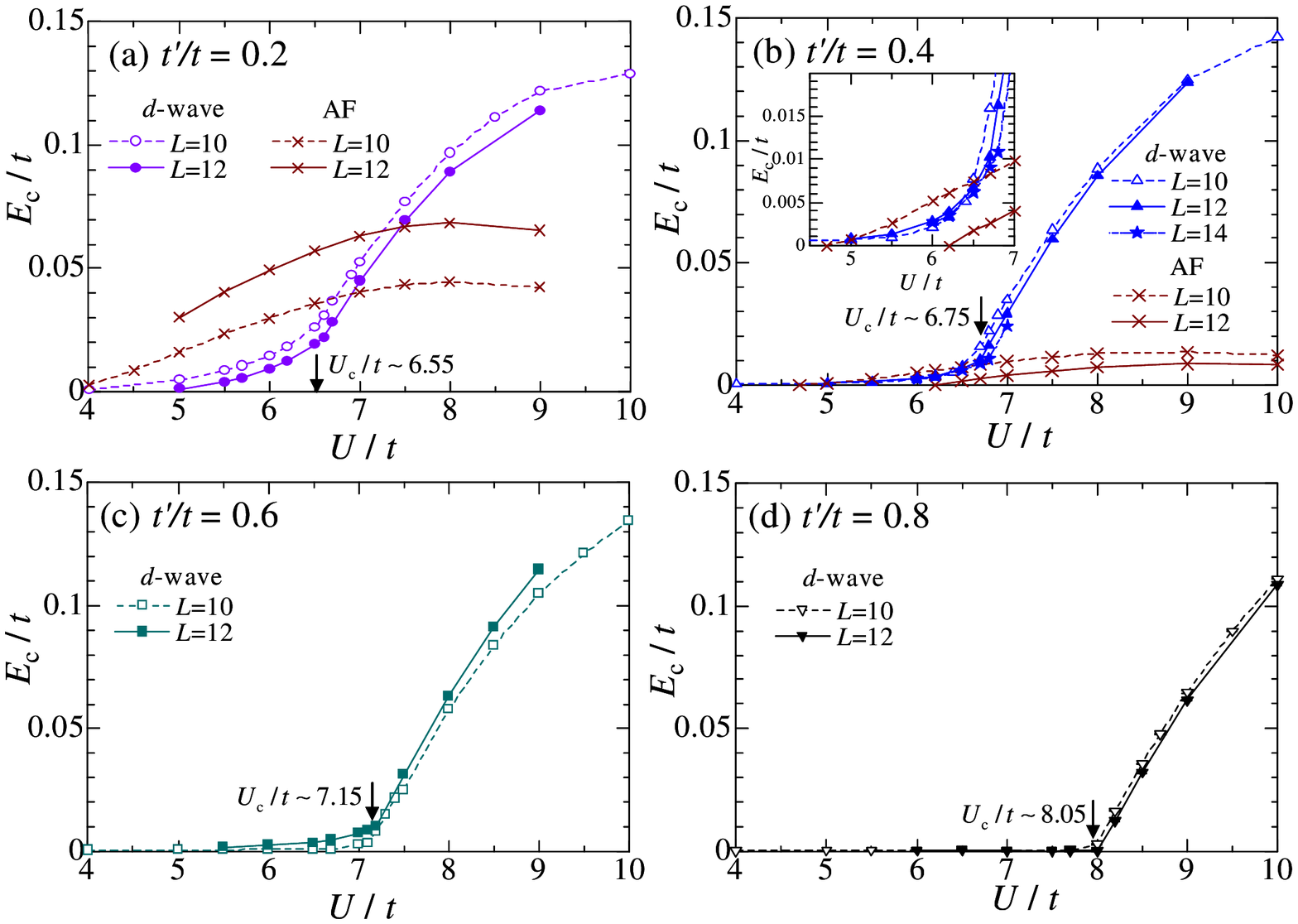}
\end{center}
\vspace{-0.5cm}
\caption{(Color online)
Condensation energies $E_{\rm c}/t$ of $d$-wave and AF states for 
four values of $t'/t$: (a) $t'/t=0.2$, (b) $0.4$, (c) $0.6$, and (d) $0.8$. 
For $t'/t=0.6$ and 0.8, $E_{\rm c}^{\rm AF}$ is zero for arbitrary 
values of $U/t$.
The arrow in each panel indicates the critical point, $U_{\rm c}/t$ 
of the Mott transition arising in the $d$-wave state. 
The values of $U_{\rm c}/t$ shown are estimated from the quantities 
treated in \S\ref{sec:mott} and \S\ref{sec:pair}.  
The inset in (b) is a close-up near the critical point. 
Systems of mainly $L=10$-$12$ are compared. 
}
\label{fig:cond}
\end{figure*}
%**********************************************************************

To consider the competition between $\Psi_Q^{\rm AF}$ and 
$\Psi_Q^d$, we take up the condensation energies defined by 
$E^{d}_{\rm c} (E^{\rm AF}_{\rm c})=E^{\rm F}-E^d(E^{\rm AF})$,
in which $E^d$, $E^{\rm AF}$, and $E^{\rm F}$ are the optimized 
variational energies per site with respect to $\Psi_Q^d$, 
$\Psi_Q^{\rm AF}$ and $\Psi_Q^{\rm F}$, respectively. 
For $t'=0$ (the regular square lattice), it is well-known 
that the ground state has an AF long-range order for arbitrary positive 
values of $U/t$ due to the complete nesting condition; this 
feature can be properly described by the present approach, 
namely $E_{\rm c}^{\rm AF}>E_{\rm c}^d>0$.\cite{YTOT} 
The maximum $E_{\rm c}^{\rm AF}$ is $0.148t$ ($U/t=8$) 
for $L=12$, and will be a little larger for $L\rightarrow\infty$. 
\cite{YTOnew}
\par

In Fig.~\ref{fig:cond}, we show $E^{d}_{\rm c}$ and 
$E^{\rm AF}_{\rm c}$ for four values of $t'/t$ ($=0.2$-$0.8$).  
As the frustration becomes stronger ($t'/t$ increases), the magnitude 
of $E^{\rm AF}_{\rm c}$ rapidly decreases, and thereby the stable 
region of the AF state shrinks. 
For $t'/t=0.2$, the AF region still remains: $U\lsim 6.7$ ($L=10$) and 
$U\lsim 7.5$ ($L=12$). 
For $t'/t=0.4$, however, the AF region becomes very narrow for $L=10$ 
($5\lsim U\lsim 6.5$), and finally vanishes for $L=12$
($E^{\rm AF}_{\rm c}<E^d_{\rm c}$), as seen in the inset of 
Fig.~\ref{fig:cond}(b). 
These indicate that at $t'/t\sim 0.4$, the ground state switches 
from the AF state to the $d$-wave state, although the system-size 
dependence of $E_{\rm c}^{\rm AF}$ is not simple near $t'/t=0.4$ 
(see $m_{\rm s}$ in Fig.~\ref{fig:pdmax}). 
For $t'/t=0.6$ and $0.8$, the AF state is never stabilized 
with respect to the normal state. 
This feature is natural at half filling, because the nesting 
of the Fermi surface monotonically becomes worse, as $t'/t$ increases. 
\par

In the following, we consider the behavior of $E^d_{\rm c}$. 
For small values of $U/t$ ($\lsim 4$), $E^d_{\rm c}$ is very small 
for arbitrary values of $t'/t$, indicating that SC is fragile, if any. 
$E^d_{\rm c}$ gradually increases for intermediate values of $U/t$: 
$4\lsim U/t\lsim 6.5$ for $t'/t=0.2$, 
$5.5\lsim U/t\lsim 6.5$ for $t'/t=0.4$, and 
$6\lsim U/t\lsim 7$ for $t'/t=0.6$. 
For $t'/t=0.8$, this gradual increase is hardly seen. 
As will be shown in \S\ref{sec:pair}, this increase corresponds to 
the stabilization due to robust SC. 
For every $t'/t$, $E^d_{\rm c}$ suddenly and markedly 
increases at $U=U_{\rm c}$, which is indicated by arrows in 
Fig.~\ref{fig:cond}.  
This increase in $E^d_{\rm c}$ certainly indicates the stabilization 
of the $d$-wave singlet state, but $\Psi_Q^d$ is not necessarily 
a SC state, as explained in \S\ref{sec:model}. 
We will pursue this behavior in the next section. 
\par

%%%%%%%%%%%%%%%%%%%%%%%%%%%%%%%%%%%%%%%%%%%%%%%%%%%
\section{Mott transitions\label{sec:mott}}
%%%%%%%%%%%%%%%%%%%%%%%%%%%%%%%%%%%%%%%%%%%%%%%%%%%
In this section, we first demonstrate that the critical behavior found 
in $\Psi_Q^d$ in \S\ref{sec:comparison} is attributed to a Mott transition 
by studying various quantities (\S\ref{sec:mott1}), and then reveal 
the properties of the transition from different points of view 
(\S\ref{sec:mott2}). 
\par

%%%%%%%%%%%%%%%%%%%%%%%%%%%%%%%%%%%%%%%%%%%%%%%%%%%
\subsection{Confirmation of Mott transitions\label{sec:mott1}}
%%%%%%%%%%%%%%%%%%%%%%%%%%%%%%%%%%%%%%%%%%%%%%%%%%%

%*****************************************************************************
% Variational parameters
%*****************************************************************************
\begin{figure*}[!t]
\begin{center}
\includegraphics[width=15cm,height=15cm]{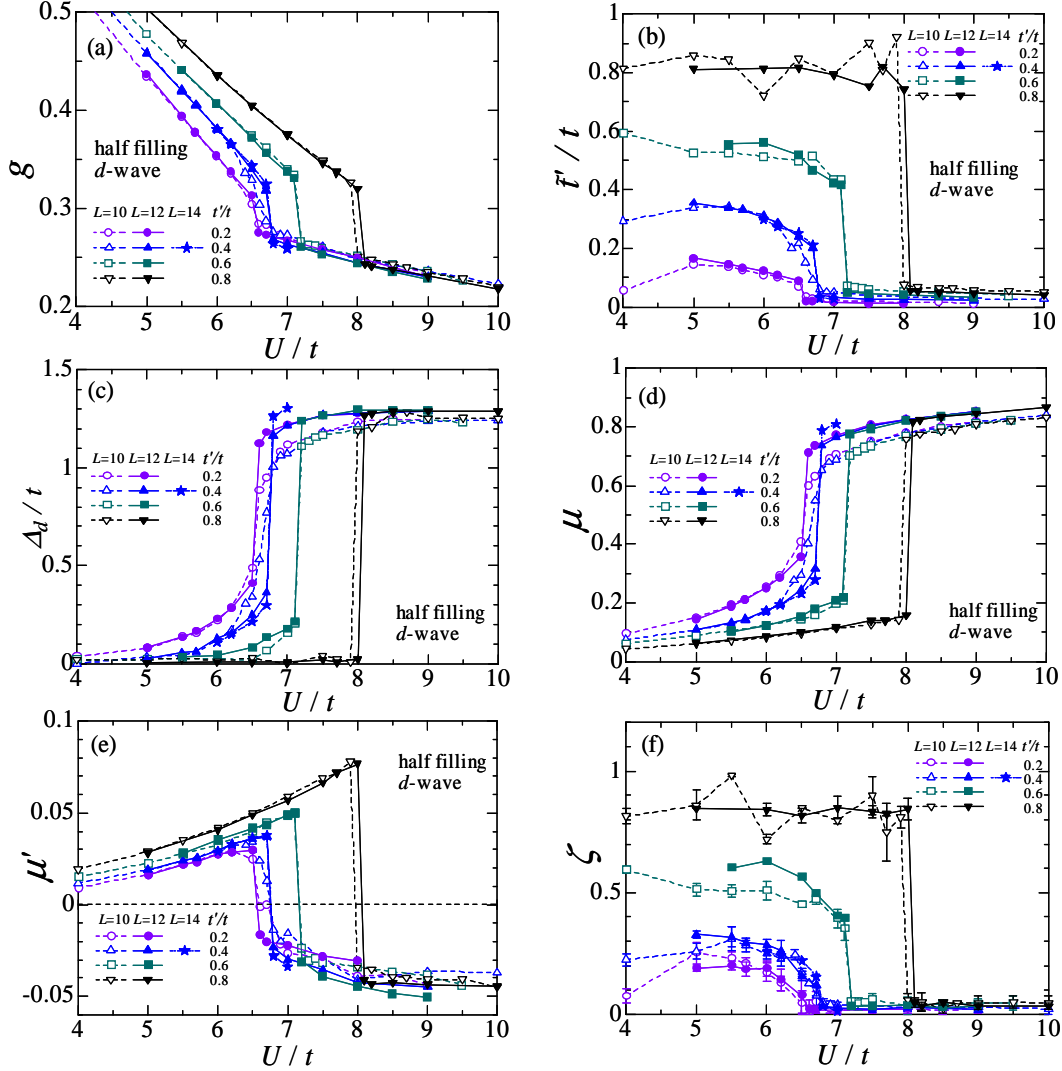}
\end{center}
\vspace{-0.5cm}
\caption{(Color online) 
Optimized values of variational parameters in $d$-wave singlet 
state, $\Psi_Q^d$, for four values of $t'/t$ as functions of $U/t$; 
(a) $g$ [onsite (Gutzwiller) factor], (b) $\tilde t'/t$ [band 
renormalization factor], (c) $\Delta_d/t$ [$d$-wave gap], (d) $\mu$ 
[doublon-holon binding factor in the direction of $t$], (e) $\mu'$ 
[the same of $t'$], (e) $\zeta$ [chemical potential]. 
The systems are of $L=10$-$14$. 
Statistical errors in $\tilde t'/t$ and in $\zeta$ increase as $U/t$ 
decreases or $t'/t$ increases, due to the discrete energy levels or 
$k$ points.
}
\label{fig:para}
\end{figure*}
%*****************************************************************************
%
First, we consider the optimized values of the variational parameters 
in $\Psi_Q^d$, which are plotted in Fig.~\ref{fig:para} for four values 
of $t'/t$.  
For each $t'/t$, all the variational parameters have discontinuities 
at the same $U/t$, which is identical to $U_{\rm c}/t$ mentioned 
in the preceding section. 
Here, $U_{\rm c}$ is determined accurately: 
$U_{\rm c}/t=6.55, 6.75, 7.15$, and 8.05 
for $t'/t=0.2$, 0.4, 0.6, and 0.8, respectively. 
These discontinuities indicate that some first order phase transition 
takes place at $U=U_{\rm c}$. 
Actually, near $U=U_{\rm c}$, $E^d$ has double minima in the parameter 
space; a hysteresis in $E^d$ is observed, as to whether we approach 
$U_{\rm c}$ from the small-$U$ side or the large-$U$ side, and particularly 
notable in the cases of large $t'/t$. 
Here, recall that the doublon-holon binding factor $\mu$ is a good index 
of a Mott transition. 
In the small-$U$ regime, $\mu$ is as small as 0.1-0.3 and almost 
independent of $L$, whereas $\mu$ is likely to approach 1 for $U>U_{\rm c}$, 
in allowing for the large $L$ dependence [Fig.~\ref{fig:para}(d)]. 
Thus, this transition is expected to be a Mott transition. 
In the following, we will confirm this expectation with various 
quantities. 
\par

\begin{figure}
\begin{center}
\includegraphics[width=7.5cm,height=9.5cm]{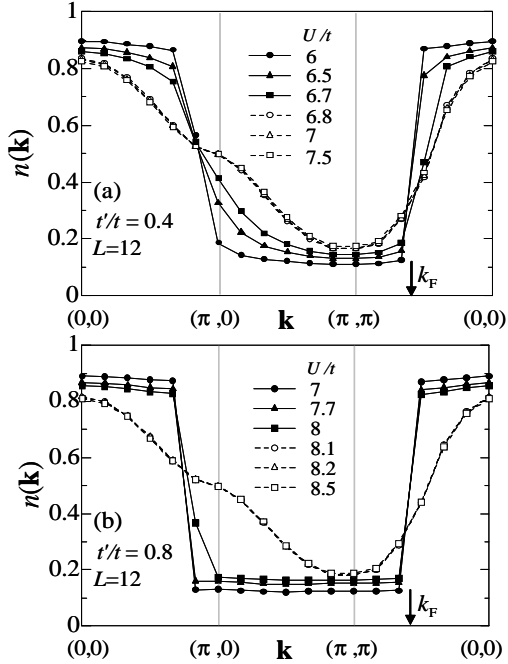}
\end{center}
\vspace{-0.5cm}
\caption{
Momentum distribution function $n({\bf k})$ of $d$-wave SC state 
for (a) $t'/t=0.4$ ($U_{\rm c}/t\sim 6.75$) 
and (b) $0.8$ ($U_{\rm c}/t\sim 8.05$). 
The solid (open) symbols denote the points of $U<U_{\rm c}$ 
($U>U_{\rm c}$). 
The arrows indicate the positions of quasi-Fermi surface in 
the node-of-gap direction (0,0)-$(\pi,\pi)$.
The systems are of $L=12$. 
}
\label{fig:mom}
\end{figure}

Let us start with the momentum distribution function 
%----------------------------------------------------------------------
\begin{equation}
n({\bf k}) = \frac{1}{2N_{\rm s}} 
\sum_{k,\sigma}\langle c_{{\bf k}\sigma}^\dag c_{{\bf k}\sigma}\rangle. 
\end{equation} 
%----------------------------------------------------------------------
%
Shown in Fig.~\ref{fig:mom} is $n({\bf k})$ measured with the optimized 
parameters along the path $(0,0)$-$(\pi,0)$-$(\pi,\pi)$-$(0,0)$ for 
$t'/t=0.4$ and $0.8$. 
Because $\Psi_Q^d$ has a node of the gap in the $(0,0)$-$(\pi,\pi)$ 
direction, metallic properties should develop in this direction; 
$n({\bf k})$ should have a discontinuity at $k_{\rm F}$. 
For both values of $t'/t$, the discontinuity at $k_{\rm F}$ is apparent 
for $U<U_{\rm c}$, whereas, for $U>U_{\rm c}$, the discontinuity, 
that is, the quasi-Fermi surface disappears. 
Thus, metallic properties are abruptly lost at $U=U_{\rm c}$ even 
in the nodal direction of the $d_{x^2-y^2}$ wave. 
\par

\begin{figure}
\begin{center}
\includegraphics[width=8.0cm,height=5.9cm]{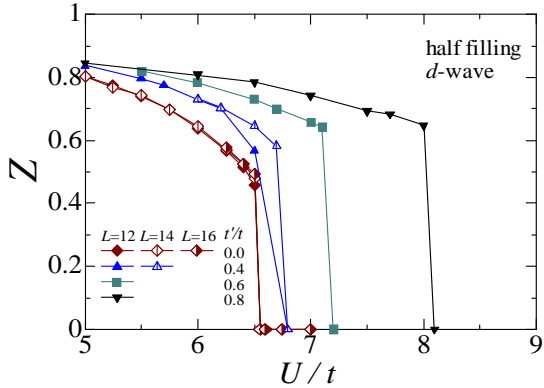}
\end{center}
\vspace{-0.5cm}
\caption{(Color online) 
Quasiparticle renormalization factor $Z$ in node-of-gap 
direction of $d$-wave singlet states for four values 
of $t'/t$ as a function of $U/t$. 
The deviation of the data for $t'/t=0.4$, $L=12$ and $U/t=6.5$ is a 
special case, and is ascribed to the discrete $k$ points of the 
system. \cite{noteZ}
}
\label{fig:ren}
\end{figure}

To see this feature quantitatively, the quasi-particle renormalization 
factor $Z$ is a suitable indicator, which roughly corresponds 
to the inverse of effective mass, unless the $k$-dependent 
renormalization of self energy is severe. 
In Fig.~\ref{fig:ren}, we plot $Z$ estimated from the magnitude of 
jump in $n({\bf k})$ at ${\bf k}={\bf k}_{\rm F}$ in the nodal direction.
\cite{noteZ} 
For all the values of $t'/t$, $Z$ decreases slowly for $U<U_{\rm c}$, 
whereas, at $U=U_{\rm c}$, $Z$ suddenly vanishes with a sizable 
discontinuity, reflecting the first-order character of the transition. 
The system-size dependence of $Z$ is weak, and the magnitude of 
$Z$ tends to increase as $L$ increases ($t'/t=0$). 
The behavior of $Z$ strongly suggests that the effective electron mass 
diverges for $U>U_{\rm c}$. 
Incidentally, the discontinuity in $Z$ at $U=U_{\rm c}$ is a feature 
that differs from those of $\Psi_Q^{\rm F}$ \cite{YokoPTP,YTOT} and 
of a dynamical mean-field theory for the hypercubic lattice,\cite{Bulla} 
in which $Z$ gradually decreases and vanishes at $U=U_{\rm c}$ 
without a jump. 
Thus, the first-order character of the transition is more conspicuous 
in $\Psi_Q^d$.
\par

%*****************************************************************************
% Momentum distribution
%*****************************************************************************
%*****************************************************************************

%*****************************************************************************
% Renormalization factor
%*****************************************************************************
%*****************************************************************************

\begin{figure}
\begin{center}
\includegraphics[width=7.5cm,height=9.5cm]{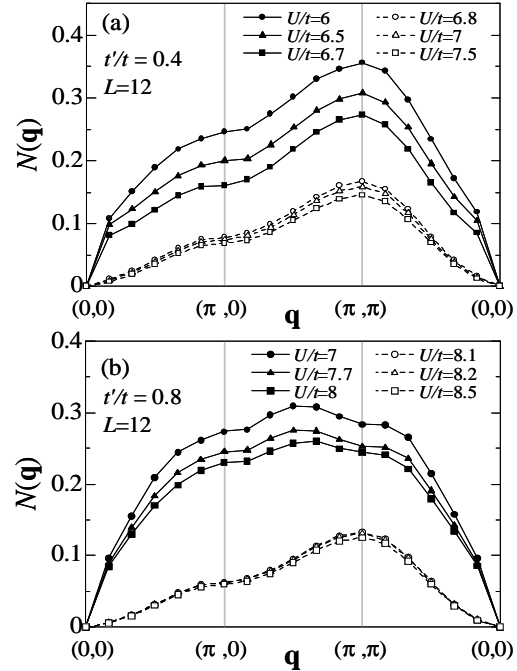}
\end{center}
\vspace{-0.5cm}
\caption{
Charge structure factor $N({\bf q})$ of $d$-wave singlet state 
for (a) $t'/t=0.4$ and (b) $0.8$. 
Solid (open) symbols indicate the data for $U<U_{\rm c}$ 
($U>U_{\rm c}$).
The systems are of $L=12$. 
}
\label{fig:charge}
\end{figure}

Next, we consider the charge structure factor 
%----------------------------------------------------------------------
\begin{equation}
N({\bf q})=\frac{1}{N_{\rm s}} 
\sum_{i,j}e^{i{\bf q}\cdot({\bf R}_i-{\bf R}_j)} 
\left\langle{N_{i} N_{j}}\right\rangle - n^2, 
\end{equation} 
%----------------------------------------------------------------------
with $N_{i} = n_{{i}\uparrow} + n_{{i}\downarrow}$. 
Within the variational theory, 
$N({\bf q})\propto |{\bf q}|$ for $|{\bf q}|\rightarrow 0$, 
if the state does not have a gap in the charge degree of freedom, 
whereas $N({\bf q})\propto {\bf q}^2$, if a charge gap opens. 
In Fig.~\ref{fig:charge}, $N({\bf q})$ is depicted for two different 
values of $t'/t$ (0.4 and 0.8) near $U=U_{\rm c}$. 
For both values of $t'/t$, $N({\bf q})$ near the $\Gamma$ point 
$(0,0)$ seems linear in $|{\bf q}|$ for $U<U_{\rm c}$ (solid symbols), 
whereas it abruptly changes to being 
roughly quadratic in $|{\bf q}|$ for $U>U_{\rm c}$ (open symbols). 
It follows that $\Psi_Q^d$ is gapless in the charge sector and is 
conductive for $U<U_{\rm c}$, but a charge gap opens for $U>U_{\rm c}$ 
and $\Psi_Q^d$ becomes insulating. 
\par

%*****************************************************************************
% Charge structure factor
%*****************************************************************************
%*****************************************************************************

\begin{figure}
\begin{center}
\includegraphics[width=8.0cm,height=6.8cm]{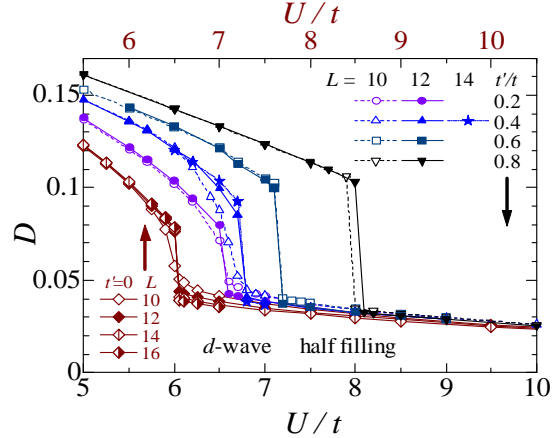}
\end{center}
\vspace{-0.5cm}
\caption{(Color online) 
Density of doublon (doubly occupied site) of $d$-wave singlet 
state as a function of $U/t$ near $U=U_{\rm c}$ for four values of 
$t'/t$. 
For $t'/t=0$, the horizontal axis is shifted by 0.5 (see upper axis)
for clarity. 
This quantity is reduced to 0.25 (0) in the limit of 
$U/t\rightarrow 0$ $(\infty)$ at half filling. 
Data for some system sizes are shown for each $t'/t$. 
}
\label{fig:doublon}
\end{figure}

As further evidence for a Mott transition, we consider the doublon 
density (density of doubly-occupied sites), 
\begin{equation}
D=\frac{1}{N_{\rm s}}\sum_i{n_{i\uparrow}n_{i\downarrow}}
=\frac{1}{N_{\rm s}}\frac{\langle{\cal H}_{\rm int}\rangle}{U}, 
\end{equation}
which is regarded as the order parameter of Mott transitions, 
\cite{Kotliar}
by analogy with the particle density in gas-liquid transitions. 
In Fig.~\ref{fig:doublon}, we plot $D$ for several values of 
$t'/t$. 
As the interaction strength $U/t$ increases, the magnitude of 
$D$ decreases almost linearly in $U/t$ for small $U/t$, but abruptly 
drops at $U_{\rm c}/t$, and then decreases slowly for 
$U>U_{\rm c}$. \cite{notesize} 
This behavior of $D$ corroborates the idea of a first-order Mott 
transition. 
\par 

%*****************************************************************************
% Density of doublon
%*****************************************************************************
%*****************************************************************************

Owing to the behavior of all the quantities studied above,
we can safely judge that a Mott transition takes place in the 
$d$-wave singlet state at $U=U_{\rm c}$ for arbitrary values of $t'/t$. 
\par

%%%%%%%%%%%%%%%%%%%%%%%%%%%%%%%%%%%%%%%%%%%%%%%%%%%
\subsection{Properties of Mott transitions\label{sec:mott2}}
%%%%%%%%%%%%%%%%%%%%%%%%%%%%%%%%%%%%%%%%%%%%%%%%%%%

For a start, we contrast this transition with the well-known theory of 
Brinkman and Rice,\cite{BR} which is based on $\Psi_{\rm G}$. 
Their theory is a milestone in the research on Mott transitions, 
and still survives as a symbolic concept of a certain class 
of transitions. 
However, it should be noted that the result of their approximation 
additionally applied, the so-called Gutzwiller approximation\cite{GA} 
(GA), has been proven correct only in infinite space dimensions. 
In accurate calculations of $\Psi_{\rm G}$, \cite{YS,YokoPTP,Metzner,Dongen} 
a metal-insulator transition is absent in finite dimensions. 
We sum up the properties of the Brinkman-Rice transition. 
For a half-filled-band Hubbard model, as $U/t$ increases toward 
$U=U_{\rm BR}=8E_0$ ($E_0$: energy for $U=0$), electrons are 
gradually localized in lattice sites one by one, and the magnetic 
susceptibility diverges as $U\rightarrow U_{\rm BR}$; this transition 
is a continuous type. 
For $U>U_{\rm BR}$, electrons are completely localized and never 
move, namely $E_{\rm kin}=E_{\rm int}=0$. 
Hence, the behavior of $|E|\sim 4t^2/U$ $(=J)$ in the insulating regime 
cannot be derived from this framework. 
\par

In contrast, in the present theory, 
the Mott transition is described not by the disappearance of charge 
carriers, but by the binding (and unbinding) of negative carriers 
(doublons) to positive carriers (holons), as explained for ${\cal P}_Q$ 
in \S\ref{sec:model} and schematically shown in Fig.~\ref{fig:d-h2}. 
In this mechanism, doublons (and holons) never vanish even for 
$U>U_{\rm c}$, as we can actually observe $g>0$ in Fig.~\ref{fig:para}(a) 
and $D>0$ in Fig.~\ref{fig:doublon}. 
The process of creating a temporary pair of a doublon and a holon 
through the so-called virtual hopping leads to the behavior proportional 
to $t^2/U$. 
\par

In this connection, a couple of studies \cite{Powell,FCZhang} 
have recently addressed the present subject, applying GA for 
{\it $d$-wave BCS states} \cite{ZGRS} to Hubbard-type models 
with extra exchange-coupling terms. 
We compare the results of these studies with those of conventional 
GA for the {\it Fermi sea} in Appendix\ref{sec:GA}. 
As long as the Mott transition is concerned, the results of GA 
for the $d$-wave BCS state have features qualitatively different 
from the present VMC results, and share the behavior with the 
conventional GA for the Fermi sea. 
\par 

Next, we consider the function of $\mu'$. 
For $U<U_{\rm c}$, $\mu'$ gradually increases, as $U/t$ increases, 
similarly to $\mu$, although $|\mu'|$ is small compared with $|\mu|$ 
[Figs.~\ref{fig:para}(d) and \ref{fig:para}(e)]. 
Thus, $\mu'$ plays an analogous role to $\mu$ in the conductive phase. 
Concerning the $t'/t$ dependence, the behavior of $\mu'$ is opposite to that 
of $\mu$, as naturally expected. 
At $U=U_{\rm c}$, however, $\mu'$ suddenly drops to a negative value, 
irrespective of $t'/t$. 
This behavior means that, in the diagonal direction, a doublon and 
a holon as two independent particles are more favorable than those 
as a bound pair in the neighbors. 
As we will demonstrate shortly, this is probably because the AF 
short-range correlation noticeably develops for $U\gsim U_{\rm c}$; 
thereby, parallel spin configurations are favored in the diagonal 
directions, and suppress the hopping in terms of $t'$. 
This marked direction dependence of $\mu({\bf r})$ is a new feature 
specific to two dimensions, in particular, to the $d$-wave singlet 
state. 
According to the exact-diagonalization calculations for one-dimensional 
systems, the behavior of $\mu(r)$ for a large $U/t$ is simple, namely, 
$\mu(r)$ is approximately proportional to $\exp(-r/r_0)$, in which 
$r_0$ is a constant. \cite{YSMott} 
\par

%*****************************************************************************
% Band width and Uc
%*****************************************************************************
\begin{figure}
\begin{center}
\includegraphics[width=6.4cm,height=4.8cm]{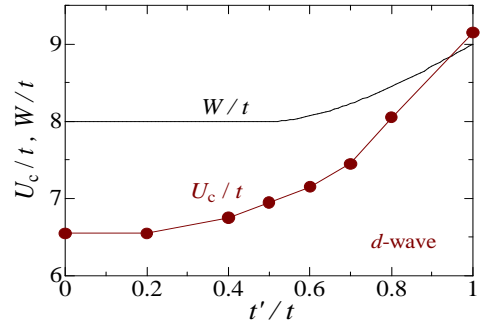}
\end{center}
\vspace{-0.5cm}
\caption{(Color online) 
Comparison between bare bandwidth $W$ and critical value 
$U_{\rm c}$ of Mott transition for $d$-wave singlet state. 
Because $U_{\rm c}/t$ is determined using finite-sized data (mainly 
$L=12$), the critical values for $L=\infty$ may somewhat shift 
to the larger side of $U/t$, especially for large values of $t'/t$.
}
\label{fig:wuc}
\end{figure}
%*****************************************************************************

In what follows, we consider the $t'/t$ dependence of this Mott 
transition in some points. 
In Fig.~\ref{fig:wuc}, the critical value $U_{\rm c}$ is compared 
with the bare bandwidth $W$. 
For the present dispersion eq.~(\ref{eq:dispersion}), $W$ is 
constant, $8t$, for $0\le t'/t\le 0.5$, and gradually increases 
as $t'/t$ increases over 0.5. 
As $W/t$ increases, the absolute value of kinetic energy increases, 
leading to a situation advantageous for the conductive state. 
Consequently, $U_{\rm c}/t$ increases with $W/t$; actually, 
$U_{\rm c}/t$ starts to increase at a smaller $t'/t$ than 
$W/t$. 
This feature is common to different Mott-transition systems. 
The ratio $U_{\rm c}/W$ has become a topic in some studies; 
both dynamical mean field calculations for infinite-dimensional 
lattices, \cite{DMFT} and VMC calculations using $\Psi_Q^{\rm F}$ 
\cite{YokoPTP,YTOnew} for $t'/t=0$ yield values of $U_{\rm c}/W$ 
somewhat larger than 1. 
In this work, however, $U_{\rm c}/W=0.81$ for $t'/t=0$, and 
increases with $t'/t$ toward 1.02 ($L=12$) for 
$t'/t=1.0$, which is relatively close to 
the result of DCA, 1.15-1.2, for $t'/t=1$. \cite{DCA}  
Conversely, this ratio obtained using PIRG is rather small, 
$U_{\rm c}/W=0.58$, even for $t'/t=1.0$. \cite{Morita}
\par

Now, we turn to the order of the Mott transition. 
Returning to Fig.~\ref{fig:para}, we notice that the discontinuity 
at $U=U_{\rm c}$ increases, as $t'/t$ increases, 
in every parameter. 
Such a tendency is common to other quantities like $Z$ (Fig.~\ref{fig:ren}) 
and $D$ (Fig.~\ref{fig:doublon}), and is also observed in the hysteresis 
or double-minimum structure in $E^d$, which becomes more conspicuous, 
as $t'/t$ increases (not shown). 
This indicates that the character of a first-order transition 
becomes prominent, as the frustration becomes stronger ($t'/t$ increases). 
This feature seems unique to $\Psi_Q^d$, and in contrast to that 
of the Mott transition in $\Psi_Q^{\rm F}$,\cite{YTOnew} as well as 
that of PIRG. \cite{Imada} 
In these cases, the effect of frustration makes the transitions more 
like those of the continuous type. 
This characteristic behavior of $\Psi_Q^d$ is probably caused by 
the prominent stability in the insulating regime, studied in the following 
and \S\ref{sec:discussions1}. 
\par

%*****************************************************************************
% Spin correlation function
%*****************************************************************************
\begin{figure}
\begin{center}
\includegraphics[width=7.5cm,height=10cm]{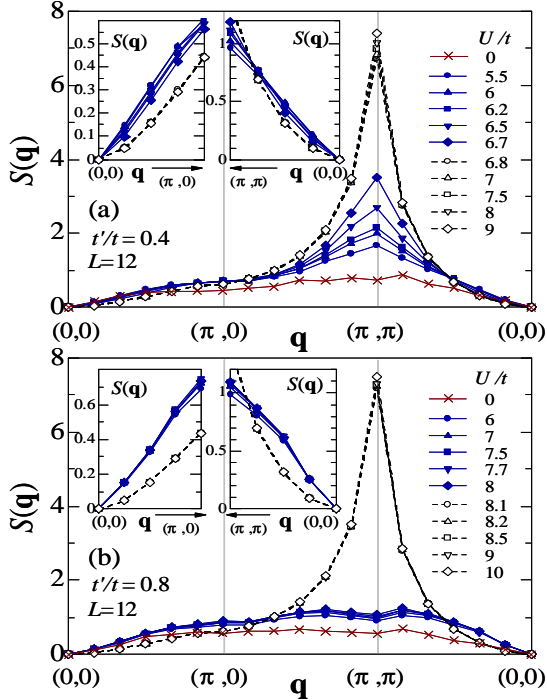}
\end{center}
\vspace{-0.5cm}
\caption{(Color online) 
Spin structure factor $S({\bf q})$ of $d$-wave singlet state along the 
same path as in Fig.~\ref{fig:mom} are compared between (a) $t'/t=0.4$ 
and (b) $0.8$. 
Solid (open) symbols denote the points belonging to the conductive 
(insulating) region. 
The insets in each panel are close-ups near the $\Gamma$ point 
$(0,0)$ in the directions of $(0,0)$-$(\pi,0)$ (left) and 
$(0,0)$-$(\pi,\pi)$ (right).
The systems are of $L=12$. 
}
\label{fig:spin}
\end{figure}
%*****************************************************************************

\begin{figure}
\begin{center}
\includegraphics[width=8.5cm,height=3cm]{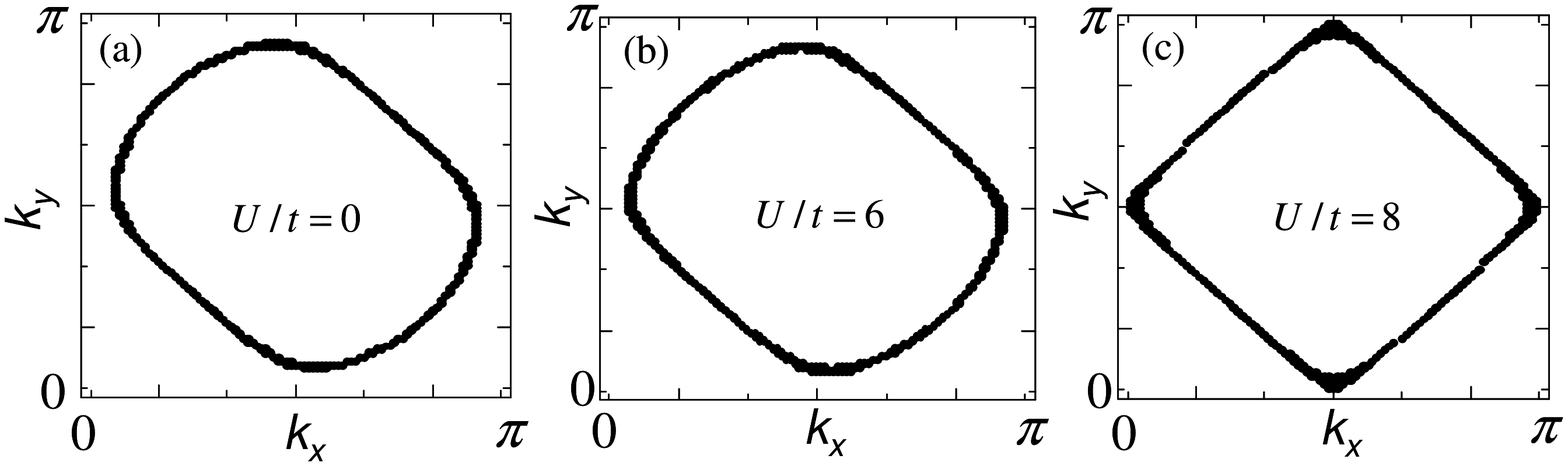}
\end{center}
\vspace{-0.5cm}
\caption{
Effective quasi-Fermi surfaces for (a) $U/t=0$, (b) $6$ (SC phase) and 
(c) $8$ (insulating phase). 
Here, the model value $t'/t$ is fixed at $0.4$, whereas the optimized 
value of $\tilde t'/t$ in the wave function is renormalized as 
$\tilde t'/t=0.31\ (0.02)$ for $U/t=6\ (8)$. 
}
\label{fig:fermi}
\end{figure}

In the remainder of this subsection, we consider the spin degree of 
freedom of the Mott insulating state described by $\Psi_Q^d$ 
for $U>U_{\rm c}$. 
Shown in Fig.~\ref{fig:spin} is the spin structure factor, 
%----------------------------------------------------------------------
\begin{equation}
S({\bf q})=\frac{1}{N_{\rm s}}\sum_{ij}{e^{i{\bf q}
\cdot({\bf R}_i-{\bf R}_j)} 
\left\langle{S_{i}^zS_{j}^z}\right\rangle},
\label{eq:sq}
\end{equation} 
%----------------------------------------------------------------------
for two values of $t'/t$. 
Leaving the behavior in the conductive regime for the next section, 
here we focus on the insulating state shown by open symbols.  
For $U>U_{\rm c}$, $S({\bf q})$ has a prominent peak at 
${\bf q}={\bf K}=(\pi,\pi)$, compared with that for $U<U_{\rm c}$; 
the AF spin correlation markedly develops in the insulating phase. 
On the other hand, the height of peaks, $S({\bf K})$, has only weak 
system-size dependence compared with $L$ as well as $L^2$: 
For example for $t'/t=0.4$ and $U/t=7$, $S({\bf K})=6.36$, 6.85, and 
7.28 for $L=10$, 12, and 14, respectively. 
In the same sense, the staggered magnetization 
\begin{equation}
m_{\rm s}=\frac{1}{N_{\rm s}}\left|\sum_i(-1)^{|{\bf R}_i|}
          \langle S_i^z\rangle\right| 
\label{eq:ms}
\end{equation}
with $S_i^z=(n_{i\uparrow}-n_{i\downarrow})/2$,\cite{metric} is 
always zero for $\Psi_Q^d$ within the statistical fluctuation. 
Thus, this insulating state has sizable short-range AF correlation, 
although a long-range order is not formed. 
The enhancement of AF correlation reflects the change in some 
quantities. 
In Fig.~\ref{fig:para}(b), where the renormalized value of the 
diagonal hopping integral is plotted versus $U/t$, we find that 
$\tilde t'/t$ suddenly drops to very small values at $U=U_{\rm c}$, 
irrespective of the model value $t'/t$. 
Therefore, the quasi-Fermi surface recovers the nesting condition 
to a considerable extent, as shown in Fig.~\ref{fig:fermi}(c). 
This feature was also reported in a study using DCA. \cite{DCA} 
Moreover, the chemical potential $\zeta$ becomes almost zero 
[Fig.~\ref{fig:para}(f)], namely, the value of the regular square 
lattice ($t'=0$), for which the nesting condition is fully satisfied. 
\par 

%*****************************************************************************
% Fermi surface
%*****************************************************************************
%*****************************************************************************

The short-range nature of the AF spin correlation affects the low-energy 
spin excitation. 
In the insets of Fig.~\ref{fig:spin}, we depict the magnifications 
of $S({\bf q})$ near the $\Gamma$ point. 
For $U>U_{\rm c}$, $S({\bf q})$ is not proportional to $|{\bf q}|$ for 
$|{\bf q}|\rightarrow 0$ but to $|{\bf q}|^2$ in both bond and diagonal 
directions and in both values of $t'/t$, indicating that this insulating 
state has a gap in the spin degree of freedom. 
\par

In \S\ref{sec:discussions1}, this insulating state will be considered 
again in light of strong coupling theories. 

%%%%%%%%%%%%%%%%%%%%%%%%%%%%%%%%%%%%%%%%%%%%%%%%%%%
\section{Superconductivity and phase diagram\label{sec:pair}}
%%%%%%%%%%%%%%%%%%%%%%%%%%%%%%%%%%%%%%%%%%%%%%%%%%%

In this section, we elucidate where, in the parameter space, 
the $d$-wave SC is dominant (\S\ref{sec:pair1}). 
Therefore, we construct a phase diagram of the present model. 
We also study the properties of the SC, emphasizing the relevance 
to the AF spin correlation (\S\ref{sec:pair2}).  
\par

%%%%%%%%%%%%%%%%%%%%%%%%%%%%%%%%%%%%%%%%%%%%%%%%%%%
\subsection{Range of robust superconductivity\label{sec:pair1}}
%%%%%%%%%%%%%%%%%%%%%%%%%%%%%%%%%%%%%%%%%%%%%%%%%%%

%*****************************************************************************
% Pair correlation function
%*****************************************************************************
\begin{figure*}[!t]
\begin{center}
\includegraphics[width=15cm,height=10cm]{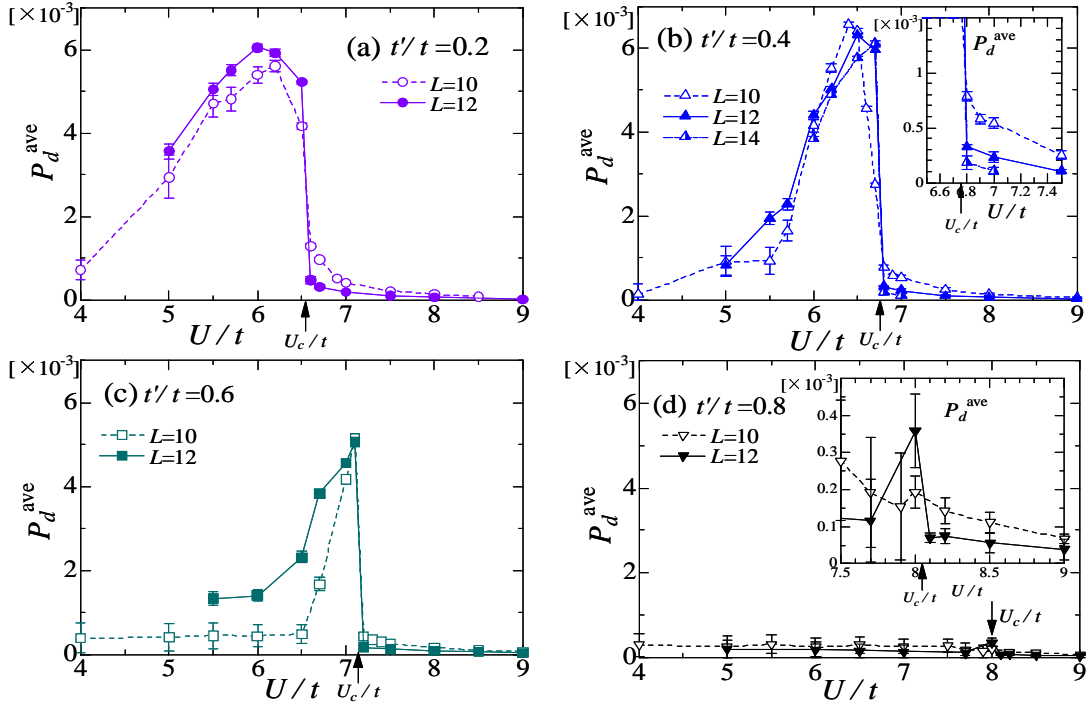}
\end{center}
\caption{(Color online)
Real-space SC correlation function for nearest-neighbor pairs 
of $d$-wave symmetry for 
(a) $t'/t=0.2$, (b) $0.4$, (c) $0.6$, and (d) $0.8$. 
Plotted are the averages of the long-distance part of $P_d({\bf r})$ 
explained in the text. 
The insets in (b) and (d) are close-ups near $U_{\rm c}/t$. 
Data of different system sizes ($L=10$-$14$) are compared. 
The same quantity for $t'/t=0.7$ is shown in Fig.~\ref{fig:liu}(b) 
in Appendix.
}
\label{fig:pair}
\end{figure*}
%*****************************************************************************

As an indicator of SC with the $d_{x^2-y^2}$ symmetry, it is 
convenient for the present approach to study the $d$-wave 
nearest-neighbor pair correlation function\cite{notePd} defined as  
%----------------------------------------------------------------------
\begin{eqnarray}
P_d({\bf r})=&&\frac{1}{4N_{\rm s}}
\sum_{i}\sum_{\tau,\tau'=\hat {\bf x},\hat {\bf y}}
(-1)^{1-\delta(\tau,\tau')}\times\qquad \nonumber\\ 
&& \left\langle{\Delta _\tau^\dag({\bf R}_i)\Delta_{\tau'}
({\bf R}_i+{\bf r})}\right\rangle, 
\label{eq:pd}
\end{eqnarray}
%----------------------------------------------------------------------
in which $\hat{\bf x}$ and $\hat{\bf y}$ denote the lattice vectors 
in the $x$- and $y$-directions, respectively, and 
$\Delta_\tau^\dag({\bf R}_i)$ is the creation operator of a
nearest-neighbor singlet expressed as 
%----------------------------------------------------------------------
\begin{equation}
\Delta_\tau^\dag({\bf R}_i)=
(c_{{i}\uparrow}^\dag c_{{i}+\tau\downarrow}^\dag+ 
 c_{{i}+\tau\uparrow}^\dag c_{{i}\downarrow}^\dag)
 /{\sqrt 2}. 
\end{equation}
%----------------------------------------------------------------------
If $P_d({\bf r})$ has a finite value for $|{\bf r}|\rightarrow\infty$, 
one can judge that an off-diagonal long-range order is realized. 
However, for finite systems, especially, in the cases with small $U/t$, 
where the correlation length is large, we have to measure long-distance 
values of $P_d({\bf r})$ appropriately. 
In this paper, we average $P_d({\bf r})$'s with ${\bf r}$ being on the 
line segments $(0,L/2)$-$(L,L/2)$ and $(L/2,0)$-$(L/2,L)$, which 
correspond to the furthermost edges from the origin, for weak 
correlation regimes\cite{noteregime}. 
For strong correlation regimes, $P_d({\bf r})$ becomes almost 
constant for $|{\bf r}|>3$;\cite{metric} we average $P_d({\bf r})$'s 
with $|{\bf r}|>3$ in these cases. 
\par

In Fig.~\ref{fig:pair}, we plot the averages of $P_d({\bf r})$ thus 
obtained ($P_d^{\rm ave}$) as a function of $U/t$ for four values of 
$t'/t$. 
The data for $t'/t=0.7$ are plotted in Fig.~\ref{fig:liu}(b).
First, let us consider features common to the cases of weak and 
intermediate frustrations ($t'/t=0.2$-$0.7$). 
For small $U/t$, the magnitude of $P_d^{\rm ave}$ is as small as 
that for the noninteracting case ($U/t=0$); here, robust SC cannot 
be expected. 
$P_d^{\rm ave}$ starts to increase appreciably at a value of $U/t$ 
($U_{\rm onset}/t$), which is different for a different $t'/t$. 
There is a peak in $P_d^{\rm ave}$ immediately below $U_{\rm c}$. 
The fact that the magnitude of $P_d^{\rm ave}$ near the peak is 
almost independent of system size indicates that a region of 
robust SC exists immediately below the Mott critical point $U_{\rm c}$. 
At $U_{\rm c}/t$, $P_d^{\rm ave}$ suddenly drops to a very small value, 
which is justifiably expected from the Mott transition. 
In the insulating regime ($U>U_{\rm c}$), as the system size $L$ 
increases, the magnitude of $P_d^{\rm ave}$ rapidly decreases as 
seen in every panel in Fig.~\ref{fig:pair}, and probably vanishes 
in the limit of $L\rightarrow\infty$. 
In the insulating regime, the statistical fluctuation in the VMC 
data is much smaller than that in the conductive regime; the 
disappearance of $P_d({\bf r})$ for $U>U_{\rm c}$ is absolutely 
certain. 
In \S\ref{sec:Liu1}, we comment on a recent VMC study,\cite{Liu} 
which has drawn contradictory conclusions from these results. 
\par 

%*****************************************************************************
% Pd_max and Sz
%*****************************************************************************
\begin{figure}
\begin{center}
\includegraphics[width=7.6cm,height=4.9cm]{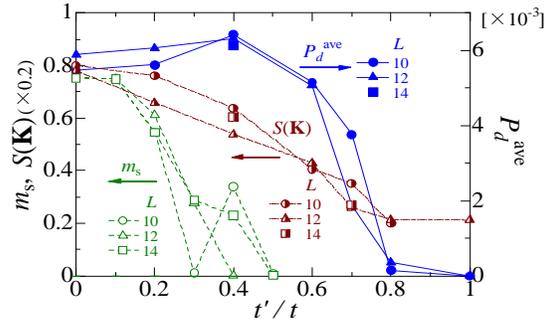}
\end{center}
\vspace{-0.5cm}
\caption{(Color online)
Comparison between the maximal values of $P_d^{\rm ave}$ (average of 
the SC correlation function) for fixed values of $t'/t$ (with $U/t$ 
changed) and $S({\bf K})$ (spin structure factor at the AF wave number) 
calculated with the same parameter of $P_d^{\rm ave}$. 
Also shown is the sublattice magnetization $m_{\rm s}$ calculated 
using $\Psi_Q^{\rm AF}$ at $U/t=6.0$. 
The scatter of data points, in particular, near the critical values 
of $t'/t$, is ascribed to finite system sizes. 
Statistical errors are much smaller. 
}
\label{fig:pdmax}
\end{figure}
%*****************************************************************************

Next, we proceed to the $t'/t$-dependence of $P_d^{\rm ave}$. 
As the strength of frustration $t'/t$ increases, the onset 
$U_{\rm onset}/t$, at which $P_d^{\rm ave}$ starts to increase appreciably, 
increases: 
$U_{\rm onset}/t\sim 4$ ($t'/t=0.2$), $\sim 5$ (0.4), $\sim 6$ (0.6), 
and $\sim 7$ (0.7). 
Because, compared with the increasing rate of $U_{\rm onset}/t$, that 
of $U_{\rm c}/t$ is slow: 6.55 (0.2), 6.75 (0.4), 7.15 (0.6), and 7.45 (0.7), 
the range in $U/t$ of the dominant SC becomes narrower, as $t'/t$ increases, 
as observed in Fig.~\ref{fig:pair}. 
In Fig.~\ref{fig:pdmax}, we depict the maximal $P_d^{\rm ave}$ 
with $t'/t$ fixed and $U/t$ varied. 
This value is almost constant for $t'/t\le 0.4$, decreases as $t'/t$ 
increases beyond $t'/t=0.4$, and finally vanishes at approximately 
$t'/t=0.8$. 
Actually for a strongly frustrated case ($t'/t=0.8$), an enhancement 
in $P_d^{\rm ave}$ is hardly seen even below $U_{\rm c}/t$ (=8.05), 
as in Fig.~\ref{fig:pair}(d), in which the magnitude of $P_d^{\rm ave}$ 
obtained for $U/t=4$-$8$ is comparable to the statistical fluctuation. 
Thus, stable SC is absent from the strongly frustrated (or nearly 
isotropic) region: $t'/t\gsim 0.8$. 
Now, recall that the AF state with a long-range order is more stable 
than the SC state for $t'/t\lsim 0.4$, as demonstrated in 
\S\ref{sec:comparison}. 
This feature is reflected in the sublattice magnetization $m_{\rm s}$ 
[eq.~(\ref{eq:ms})] for $\Psi_Q^{\rm AF}$, as shown in Fig.~\ref{fig:pdmax}. 
Thus, we estimate the range of stable SC as $0.4\lsim t'/t\lsim 0.7$. 
\par

Incidentally, as shown in the insets of Fig.~\ref{fig:spin}(a), 
$S({\bf q})$ tends to be proportional to $|{\bf q}|^2$ for 
$|{\bf q}|\rightarrow 0$ as $U/t$ increases ($U<U_{\rm c}$), 
indicating a spin gap is apt to open as SC develops. 
In contrast, in the insets of Fig.~\ref{fig:spin}(b), $S({\bf q})$ remains 
linear in $|{\bf q}|$, even if $U/t$ increases; $\Psi_Q^d$ remains 
metallic for $t'/t=0.8$, even if $U$ approaches $U_{\rm c}$. 
\par

The present results for SC are consistent, even quantitatively, with 
those of FLEX. \cite{Kino,Kondo}
According to Kino and Kontani, the AF order defeats SC for $t'/t<0.4$, 
whereas the region of finite $T_{\rm c}$ of $d_{x^2-y^2}$-wave SC 
appears for $0.4<t'/t<0.8$ with the highest values taken for 
$t'/t=0.5$-$0.6$. 
Finally, $T_{\rm c}$ of SC disappears for $t'/t>0.8$. 
Furthermore, the range of SC in $U/t$ displays a tendency similar to 
that of the present study. 
\par

Having discussed the range of SC in the $t'$-$U$ space and its 
stability, let us now reconsider how various quantities behave in this 
SC region. 
First, concerning the condensation energy shown in Fig.~\ref{fig:cond}, 
we are now aware that the gradual increase in $E_{\rm c}$ for 
$U_{\rm onset}\lsim U<U_{\rm c}$ derives from SC, whereas the sizable 
gain for $U>U_{\rm c}$ is not due to SC but to the $d$-wave singlet 
spin state. 
Similar gradual changes can be seen in the variational parameters 
(Fig.~\ref{fig:para}); in particular, the increase in $\Delta_d$ 
for $U_{\rm onset}\lsim U<U_{\rm c}$ directly corresponds to the 
development of SC [Fig.~\ref{fig:para}(c)]. 
The slow decrease in $\tilde t'/t$ [Fig.~\ref{fig:para}(b)] means that 
the degree of renormalization of $\tilde\varepsilon_{\bf k}$ is low 
in the SC state; the quasi-Fermi surface is not very different from that 
for $U/t=0$ as in Figs.~\ref{fig:fermi}(a) and (b). 
This tendency is again consistent with the result of FLEX. \cite{Kino}
Note that such a gradual change never appears in any quantity 
in the case of $t'/t=0.8$, where robust SC is absent. 
\par

%%%%%%%%%%%%%%%%%%%%%%%%%%%%%%%%%%%%%%%%%%%%%%%%%%%
\subsection{Properties of superconductivity\label{sec:pair2}}
%%%%%%%%%%%%%%%%%%%%%%%%%%%%%%%%%%%%%%%%%%%%%%%%%%%

Because the SC develops as soon as the AF long-range order disappears, 
the AF correlation must be important for the mechanism of SC. 
We hence begin with the spin structure factor $S({\bf q})$ 
[eq.~(\ref{eq:sq})], returning to Fig.~\ref{fig:spin}. 
For $U/t=0$, $S({\bf q})$ does not have a peak but rather a dip 
at the AF wave number ${\bf K}=(\pi,\pi)$ for both $t'/t=0.4$ and 
0.8, due to the frustration. 
In the weakly frustrated case ($t'/t=0.4$), in which robust SC appears, 
$S({\bf K})$ steadily grows 
as $U/t$ increases, and ${\bf K}$ becomes a characteristic wave number 
even in the conductive regime, $U<U_{\rm c}$ [solid symbols in 
Fig.~\ref{fig:spin}(a)]. 
In contrast, in the strongly frustrated case ($t'/t=0.8$), in 
which the SC correlation does not develop, although $S({\bf q})$ 
somewhat increases as a whole for $U<U_{\rm c}$ [solid symbols in 
Fig.~\ref{fig:spin}(b)], preserving the shape for $U/t=0$, 
yet $S({\bf q})$ does not exhibit a special enhancement 
at ${\bf q}={\bf K}$, but has maxima at incommensurate wave numbers. 
To check this point, we plot, in Fig.~\ref{fig:pdmax}, $S({\bf K})$ 
for $\Psi_Q^d$ giving the maximal $P_d^{\rm ave}$; when 
$S({\bf K})$ decreases, the maximal $P_d^{\rm ave}$ simultaneously 
decreases. 
We have confirmed it for a wide range of model parameters that 
whenever $P_d^{\rm ave}$ is appreciably enhanced, $S({\bf q})$ 
has an evident peak at ${\bf q}={\bf K}$. 
From these discussions, we can conclude that SC in this model is 
induced by the AF spin correlation. 
Thus, we can understand that the shape of the quasi-Fermi surface of 
$\Psi_Q^d$ tends to be more square-lattice-like ($\tilde t'/t'<1$), 
as mentioned earlier, so as to place the nesting vector {\bf K} on hot spots 
having a larger density of state. 
It is probable that the mechanism of SC proposed here is basically 
the same as those of the weak-coupling theories, \cite{Kino,Kondo,Yamada}
and is also common to that of high-$T_{\rm c}$ cuprates at half filling, 
if any, which have quite similar lattice structure 
(see Fig.~\ref{fig:model}). \cite{YTOnew} 
\par

Here, we summarize the features of this SC in light of the AF correlation. 
(1) For small values of $U/t$, because magnetic correlations do not 
fully develop, robust SC does not appear. 
(2) For small values of $t'/t$, although the AF correlation develops 
as $U/t$ increases, an AF long-range order defeats SC due to the 
nesting condition. 
(3) For moderate values of $t'/t$ as well as $U/t$, the AF correlation 
is not strong enough to form an AF long-range order, but the short-range 
part survives. 
Thereby, stable $d$-wave SC is realized. 
(4) For large values of $t'/t$, even short-range AF correlation decays
due to severe frustration. 
Thus, SC is not likely to arise. 
(5) For considerably large values of $U/t$, a Mott insulating state 
suppresses SC.
\par

%*****************************************************************************
% Phase diagram
%*****************************************************************************
\begin{figure}
\begin{center}
\includegraphics[width=8cm,height=5.5cm]{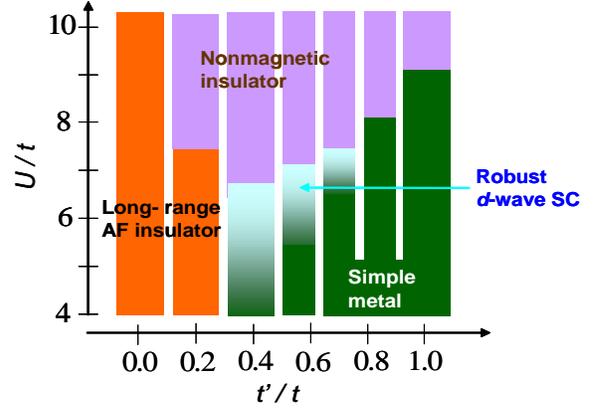}
\end{center}
\vspace{-0.5cm}
\caption{(Color online) 
Phase diagram in the $t'$-$U$ plane constructed from the present 
VMC results. 
}
\label{fig:phase}
\end{figure}
%*****************************************************************************

On the basis of the results mentioned upto this point, we constructed 
a phase diagram in the model space spanned by $t'/t$ and $U/t$ 
(Fig.~\ref{fig:phase}).  
Robust $d$-wave SC appears in the highlighted area 
($0.4\lsim t'/t\lsim 0.7$, $U_{\rm onset}\lsim U<U_{\rm c}$), which 
is in contact with both AF and Mott insulating areas, and 
continues to the metallic area in the small-$U/t$ or large-$t'/t$ sides, 
where the SC correlation is very weak. 
The conductive (metallic or SC) and nonmagnetic insulating areas 
are divided by a first-order Mott transition described within the 
$d$-wave singlet state. 
The boundary between the areas of long-range AF insulator and 
the nonmagnetic insulator (or superconductor) is determined by 
comparing the energies of $\Psi_Q^{\rm AF}$ and $\Psi_Q^d$. 
\par

%*****************************************************************************
% Energy component
%*****************************************************************************
\begin{figure}
\begin{center}
\includegraphics[width=7.5cm,height=9.5cm]{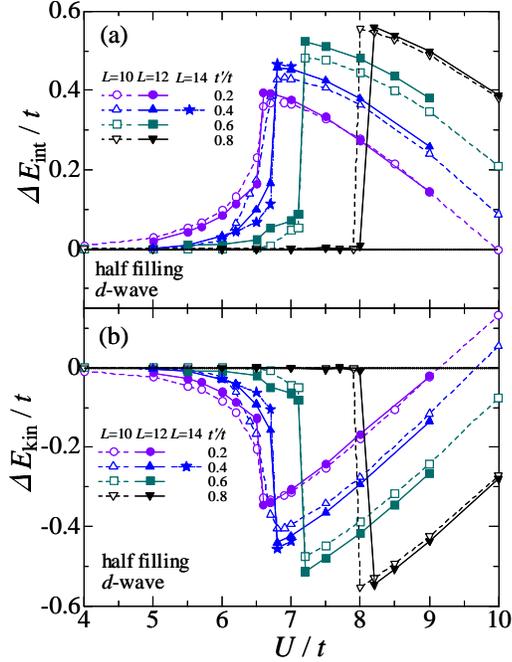}
\end{center}
\caption{(Color online) 
Differences in (a) interaction and (b) kinetic energies between 
normal and $d$-wave states for $t'/t=0.2$, $0.4$, $0.6$, and $0.8$. 
}
\label{fig:com}
\end{figure}
%*****************************************************************************

Finally, we take up another weak-coupling feature of this SC. 
Shown in Fig.~\ref{fig:com} are the components of condensation 
energy, that is, the differences in kinetic and interaction energies 
between $\Psi_Q^d$ and $\Psi_Q^{\rm F}$ given as 
\begin{eqnarray}
\Delta E^d_{\rm kin}&=&E^{\rm F}_{\rm kin}-E^d_{\rm kin}, \\
\Delta E^d_{\rm int}&=&E^{\rm F}_{\rm int}-E^d_{\rm int}, 
\end{eqnarray} 
in which $\Delta E^d_{\rm kin}+\Delta E^d_{\rm int}=E_{\rm c}^d\ (\ge 0)$. 
In the SC regime ($U<U_{\rm c}$), $\Delta E^d_{\rm int}$ 
($\Delta E^d_{\rm kin}$) is always positive (negative), regardless of 
$t'/t$. 
This indicates that the SC transition is induced by the gain in 
interaction energy at the cost of kinetic energy. 
This feature smoothly continues to the weak-coupling limit 
($U/t\rightarrow 0$), and common to conventional BCS superconductors. 
Although the component of energy gain switches to the kinetic energy 
for $U\gsim 10t$, as seen in Fig.~\ref{fig:com}, yet a SC phase is not 
realized for $U>U_{\rm c}$; the SC due to the kinetic-energy gain 
is not realized here, in contrast to the doped cases. \cite{YTOT}
\par 

%%%%%%%%%%%%%%%%%%%%%%%%%%%%%%%%%%%%%%%%%%%%%%%%%%%%%%%%%%%%%%%%%%%%%%%%%
\section{Further Discussions\label{sec:discussions}}
%%%%%%%%%%%%%%%%%%%%%%%%%%%%%%%%%%%%%%%%%%%%%%%%%%%%%%%%%%%%%%%%%%%%%%%%%
In \S\ref{sec:discussions2}, we compare the results obtained in this work 
with those of related experiments on $\kappa$-ET salts. 
In \S\ref{sec:discussions1}, as a continuation of \S\ref{sec:mott2}, 
we analyze $\Psi_Q^d$ for $U>U_{\rm c}$ in light of strong-coupling 
theories \cite{WatatJ}, and propose a couple of improvements. 
\par

%%%%%%%%%%%%%%%%%%%%%%%%%%%%%%%%%%%%%%%%%%%%%%%%%%%%%%%%%%%%%%%%%%%%%%%%%
\subsection{Comparisons with experiments \label{sec:discussions2}}
%%%%%%%%%%%%%%%%%%%%%%%%%%%%%%%%%%%%%%%%%%%%%%%%%%%%%%%%%%%%%%%%%%%%%%%%%
%
First, let us think of SC. 
If the present results for SC is applicable to $\kappa$-ET salts, the 
SC arising in them should be induced by AF correlation, which 
becomes weak as $t'/t$ increases (Fig.~\ref{fig:pdmax}). 
Actually, in a class of $\kappa$-ET salts, a compound with weaker 
frustration (smaller $t'/t$) tends to have higher $T_{\rm c}$: 
for X=${\rm Cu{N(CN)_2}Br}$ ($t'/t=0.68$), 
${\rm Cu(NCS)_2}$ (0.84), and ${\rm Cu_2(CN)_3}$ (1.06), $T_{\rm c}=11.6$, 
$10.4$, and $3.9$K (under pressure of 0.4GPa), respectively. 
Note, however, that, according to our result, robust SC does not 
arise for $t'/t\gsim 0.8$; this fact seemingly contradicts 
the experiments on $\kappa$-ET salts, particularly of the compound 
X=${\rm Cu_2(CN)_3}$. 
A point we must not overlook here is the effect of pressure. 
The compound X=${\rm Cu_2(CN)_3}$ has an almost regular triangular 
structure ($t'=t$) under ambient pressure, which fact is also confirmed 
by a high-temperature-expansion study. \cite{highTE}
However, this compound does not exhibit SC, unless a pressure of 
0.4-0.6GPa is applied. \cite{Kurosaki} 
It is well-known that applied pressure generally enhances the bandwidth 
(reduce $U/t$). 
It is reasonable to expect that $t'/t$, in addition to $U/t$, 
should be greatly varied by pressure, \cite{CuGeO3} 
if we allow for the fact that organic compounds are soft and anisotropic, 
and the values of $t$ and $t'$ are sensitive to bond angle. 
Actually, a recent NMR-$T_1$ experiment for X=${\rm Cu_2(CN)_3}$ 
\cite{Kurosaki} has suggested the enhancement of AF fluctuation under 
pressure; in this case, it is likely that $t'/t$ is 
sufficiently reduced to induce SC. 
Although the pressure effect on $t'/t$ is quite an important problem, 
it has not yet been established whether $t'/t$ is 
reduced \cite{tpRahal} or enhanced \cite{tpCampos} by pressure. 
\par

Next, we touch on Mott (SC-nonmagnetic insulator) transition. 
The existence of a first-order Mott transition is consistent with 
the experiment on temperature-dependent magnetic susceptibility 
and the nuclear spin-lattice relaxation rate of 
$\kappa$-(ET)$_2$Cu$_2$(CN)$_3$. \cite{Shimizu}
In this compound, however, the ground state in the insulating phase 
has a low-energy spin excitation; the spin gap will be very small, 
if it exists. 
This point is not necessarily consistent with the present result. 
Furthermore, the insulating states of most $\kappa$-ET salts exhibit 
a long-range AF order, although they are considered strongly frustrated 
with $t'/t\gsim 0.6$.
Even if we take account of the fact that actual materials possess 
the effect of three dimensions, the trial function in the Mott 
insulating regime leaves room for improvement with regard to 
the spin correlation, as will be mentioned in the next subsection.  
In relation to the insulating state, it is also an interesting future 
problem how the 120 degree N\'eel order, which is realized for 
$U/t\rightarrow\infty$ and $t'/t\sim 1$, depends on $U/t$ and 
$t'/t$. \cite{120degree}
\par

%%%%%%%%%%%%%%%%%%%%%%%%%%%%%%%%%%%%%%%%%%%%%%%%%%%%%%%%%%%%%%%%%%%%%%%%%
\subsection{$\Psi_Q^d$ as an RVB state \label{sec:discussions1}}
%%%%%%%%%%%%%%%%%%%%%%%%%%%%%%%%%%%%%%%%%%%%%%%%%%%%%%%%%%%%%%%%%%%%%%%%%
Although $\Psi_Q^d$ has succeeded in describing a Mott transition 
for an intermediate coupling strength, some adjustments in the spin 
correlation seem to be needed for the insulating regime ($U>U_{\rm c}$), 
where a strong-correlation scheme is convenient. 
\par

As a previous study for the square lattice showed, \cite{YTOT} 
the strong-coupling regime ($U>U_{\rm c}$ for half filling) of the 
Hubbard model shares the properties with $t$-$J$-type models; 
the physics adiabatically connected to the limit of $U/t\rightarrow\infty$. 
This picture also holds in the insulating state here, because the 
energies are scaled with $t^2/U$, and various quantities obtained 
with $\Psi_Q^d$ changes very slowly as a function of $U/t$ and $t'/t$, 
as observed in 
$n({\bf k})$ [Fig.~\ref{fig:mom}], $N({\bf q})$ [Fig.~\ref{fig:charge}], 
$D$ [Fig.~\ref{fig:doublon}] and $S({\bf q})$ [Fig.~\ref{fig:spin}]. 
In this sense, $\Psi_Q^d$ is substantially equivalent to Anderson's 
RVB state. \cite{Anderson} 
\par

For $U/t\rightarrow\infty$, the present Hubbard model eq.~(\ref{eq:model}) 
is reduced to the $J$-$J'$ Heisenberg model with the same connectivity. 
Some approximations \cite{J-J'} to this spin model concluded that 
the N\'eel order persists up to $J'/J=0.6$-$0.7$ ($t'/t=0.77$-0.84) 
beyond the classical value $J'/J=0.5$ ($t'/t=0.71$). 
In contrast, in this study, the phase boundary between the AF state 
and the insulating state of $\Psi_Q^d$ for $U\rightarrow\infty$
is situated at $t'/t<0.2$, as deduced from Fig. \ref{fig:cond}(a). 
Thus, the two approaches quantitatively give different estimates. 
A possible origin of this discrepancy is that, despite the marked 
AF short-range correlation, $\Psi_Q^d$ does not have a seed of AF 
longer-range correlations. 
Also, the fact that the renormalization of $t'/t$ is not introduced 
into $\Psi_Q^{\rm AF}$ possibly affects the discrepancy. 
As an improvement in both $\Psi_Q^d$ and $\Psi_Q^{\rm AF}$, it is 
very important to study a wave function that can possess both 
$d$-wave pairing and AF order. \cite{Himeda} 
This will expand the area of AF insulator to some extent 
(see Fig.~\ref{fig:phase}). 
\par

Nonetheless, note that $\Psi_Q^d$ has an extremely low energy as an 
insulating state; it readily defeats the AF state even with weak 
frustration, as in Fig.~\ref{fig:cond}. 
In this connection, we need to remember Liang \etal's study \cite{Liang}, 
in which the Heisenberg model on the square lattice is studied with 
intelligible wave functions, which are composed of a product of 
singlets with a bond-length distribution as a variational parameter. 
They showed that the wave functions including long singlet bonds 
exhibit an AF long-range order with very low energies; the wave functions 
composed only of short-bond singlets do not have AF long-range order 
but have very close energies to the former. 
Thus, the stability of the wave function is little influenced by 
whether the spin correlation is short-ranged or long-ranged. 
There is no reason to suppose that this situation does not hold 
for frustrated cases.
Because Liang \etal's wave functions are not tractable for $t'/t\ne 0$, 
we should take an alternative approach, for example, refine a class 
of projected singlet wave functions by adjusting the gap parameters. 
They are expressed as the same form as eq.~(\ref{eq:singlet}) 
with $\varphi_{\bf k}$, which generally differs from that of the BCS type, 
eq.~(\ref{eq:BCSDelta}), but is tractable in the present scheme. 
\par

%%%%%%%%%%%%%%%%%%%%%%%%%%%%%%%%%%%%%%%%%%%%%%%%%%%%%%%%%%%%%%%%%%%%%%%%%
\section{Summary\label{sec:summary}}
%%%%%%%%%%%%%%%%%%%%%%%%%%%%%%%%%%%%%%%%%%%%%%%%%%%%%%%%%%%%%%%%%%%%%%%%%

We have studied a Hubbard model on an anisotropic triangular lattice 
at half filling by an optimization VMC method with adequate precision, 
taking account of the SC and Mott transition arising in $\kappa$-ET salts.
In the trial functions---normal, AF, and $d$-wave singlet states, 
we introduce intersite correlation factors that control the binding 
between a doublon and a holon. 
We have succeeded in describing the $d$-wave SC and a Mott transition 
simultaneously in a sole approach, and in accounting for the behavior 
of a series of $\kappa$-ET salts broadly. 
We itemize our main results: 
\par

(1) Within the $d$-wave singlet state, a first-order Mott 
(conductor-to-nonmagnetic insulator) transition takes place at $U_{\rm c}$ 
approximately of the bandwidth for arbitrary values of $t'/t$. 
The critical value $U_{\rm c}/t$ gradually increases as band 
width (or $t'/t$) increases. 
This transition is not directly related to a magnetic order. 
\par

(2) The nonmagnetic insulating state ($d$-wave singlet state for 
$U>U_{\rm c}$) has a considerably low energy and a strong short-range 
AF correlation. 
This state has a spin gap, as well as a charge gap. 
\par

(3) The AF long-range order prevailing in the weakly frustrated 
cases ($t'/t\lsim 0.4$) is rapidly destabilized as $t'/t$ increases, 
and yields to the $d$-wave singlet state (SC or nonmagnetic insulating). 
\par

(4) SC with the $d_{x^2-y^2}$-wave symmetry appears for moderate values 
of $U/t$ ($\sim 6$) and $t'/t$ ($0.4\lsim t'/t\lsim 0.7$), whose area 
is adjacent to both areas of an AF long-range order and a Mott insulator. 
The phase diagram obtained in this study is shown in Fig.~\ref{fig:phase}. 
\par

(5) SC pairs are formed with the aid of the short-range AF spin 
correlation, which is weakened by the frustration and vanishes 
for $t'/t\gsim 0.8$. 
The SC transition is induced by the gain in interaction energy; 
this mechanism is identical to that in the weak correlation limit 
as well as that of conventional BCS superconductors. 
\par

In this paper, we have focused on the $d_{x^2-y^2}$-wave symmetry, 
which is established for $t'/t=0$. \cite{Gros} 
It has been reported, however, that the symmetry of pairing switches 
to $d$+$(i)d$-type symmetries at $t'/t=1$. \cite{Moriya,WataCo,Liu}
It is hence worthwhile to check other pairing symmetries for 
large values of $t'/t$. 
Moreover, it is interesting to check with VMC how the features revealed 
in the single-band Hubbard model are altered in more realistic multiband 
models, \cite{Schmalian} especially in strong and intermediate 
correlation regimes. 
In connection with pairing symmetry, we finally comment that 
it is interesting to study the phase sensitive property, $e.g.$, 
the tunneling effect and Josephson effect, in unconventional 
superconductor near a Mott transition. \cite{Kashiwaya,Ichimura}
\par 

%-----------------------------------------------------------------------
% acknowledgments
%-----------------------------------------------------------------------

\begin{acknowledgments}
The authors thank M.~Ogata for useful discussions. 
They appreciate the discussions with and comments of H.~Fukuyama, 
N.~Nagaosa, K.~Kanoda, K.~Kuroki, and H.~Kontani. 
This work is partly supported by Grant-in-Aids from 
the Ministry of Education, Culture, Sports, Science and Technology, 
by NAREGI Nanoscience Project of the same Ministry,  
by the Supercomputer Center, ISSP, University of Tokyo, 
and by a Grant-in-Aid for the 21st Century COE 
``Frontiers of Computational Science". 
\end{acknowledgments} 

%-----------------------------------------------------------------------

\appendix
%%%%%%%%%%%%%%%%%%%%%%%%%%%%%%%%%%%%%%%%%%%%%%%%%%%%%%%%%%%%%%%%%%%%%%%%%
\section{Gutzwiller approximation for $d$-wave state
\label{sec:GA}}
%%%%%%%%%%%%%%%%%%%%%%%%%%%%%%%%%%%%%%%%%%%%%%%%%%%%%%%%%%%%%%%%%%%%%%%%%
%
In this Appendix, we compare the results of conductor-insulator 
transitions in recent studies \cite{Powell,FCZhang} using GA for 
{\it the $d$-wave BCS states} \cite{ZGRS} with the previous ones 
using GA for {\it the Fermi sea}. 
Two groups have recently addressed the present subject by applying 
GA for $d$-wave BCS states to Hubbard-type models with extra 
exchange-coupling terms.
These extra terms are added for technical reasons, and seem to make 
no essential difference. 
Powell and McKenzie \cite{Powell} studied a Mott transition and 
SC without allowing for AF. 
They found a first-order SC-insulator transition at $U/t$ somewhat 
larger than the bandwidth.
In the insulating regime, however, the kinetic energy vanishes, 
indicating this transition belongs to the Brinkman-Rice type. 
The AF effect is incorporated into GA for 
the $d$-wave BCS state by Gan \etal\cite{FCZhang}, who found
a SC-AF insulator transition at $U/t$ somewhat smaller than the 
bandwidth. 
In the SC regime, the SC correlation and renormalization of 
$g$ develop as $U/t$ increases without sublattice magnetization, 
whereas in the AF insulating regime, there appears a finite sublattice 
magnetization but $g$ is reduced to the value at $U/t=0$, namely 
$g=1$. 
In fact, this result is closely parallel to that of the GA for 
the {\it Fermi sea} into which an AF effect is introduced, as 
in the following. 
\par

About three decades ago, Ogawa \etal\ \cite{OKM} 
introduced the AF effect into the GA formula for the Fermi sea, using 
${\cal P}_{\rm G}\Phi_{\rm AF}$, with a cubic-type lattice in mind. 
The result obtained exhibits a transition; for $0<U<U_{\rm AF}$, 
the result coincides with the ordinary (or paramagnetic) GA formula 
($\Delta_{\rm AF}=0$, $g<1$), whereas for $U>U_{\rm AF}$, the result 
is switched to the mean-field solution of AF insulator 
($\Delta_{\rm AF}>0$, $g=1$). 
Here, $U_{\rm AF}$ denotes $U$ at which the energy curve 
of the GA formula intersects with that of the AF mean-field approximation. 
Thus, the transition at $U=U_{\rm AF}$ is a first-order AF transition 
which does not correspond to the Mott transition we considered in this paper, 
but is basically identical to those derived by the mean-field theories
for $t'\ne 0$. 
\cite{KondoHF} 
Later, it is shown by a VMC study \cite{YSAF} for 
${\cal P}_{\rm G}\Phi_{\rm AF}$ that this conclusion is again incorrect; 
the true variational state has an AF gap and a finite renormalization 
by $g$ ($\Delta_{\rm AF}>0$, $g<1$) simultaneously for arbitrary values 
of $U (>0)$, and does not exhibit a transition for $U>0$. 
\par

Thus, as long as a Mott transition is concerned, the results of GA 
for the $d$-wave BCS state \cite{Powell,FCZhang} share the features 
with the conventional GA for the Fermi sea. \cite{BR,OKM} 
We should be sufficiently careful to use GA to study a Mott transition. 
Nevertheless, it is fair to mention that the results of GA with respect 
to SC are often consistent with the present ones shown in \S\ref{sec:pair}. 
\par

%%%%%%%%%%%%%%%%%%%%%%%%%%%%%%%%%%%%%%%%%%%%%%%%%%%%%%%%%%%%%%%%%%%%%%%%%
\section{Mott transition by doublon-holon binding factors
\label{sec:liu}}
%%%%%%%%%%%%%%%%%%%%%%%%%%%%%%%%%%%%%%%%%%%%%%%%%%%%%%%%%%%%%%%%%%%%%%%%%

Up to this time, a Mott transition has been repeatedly studied using 
the doublon-holon binding factor ${\cal P}_Q$ and its minor
variations. 
However, the course of study has not been straight but a little 
confusing. 
In this Appendix, we first recapitulate this subject 
(\S\ref{sec:exp}), 
and then make comments on misunderstandings based on 
the above confusion in a recent VMC study\cite{Liu} (\S\ref{sec:Liu1}). 
\par

%-------
\subsection{Mott transition by ${\cal P}_Q$\label{sec:exp}}
%-------

For simplicity, let us consider the case of $t'=0$ ($\mu'=0$), which 
makes no substantial difference for finite $t'$. 
Because $\mu$ in eq.~(\ref{eq:PQ}) is close to 1 
near a Mott transition, the wave function $\Psi_Q$ eq.~(\ref{eq:wf}) 
is expanded with respect to $1-\mu$ ($=m$) as 
\begin{equation}
\Psi_Q=\Psi_Q^{(0)}+m\Psi_Q^{(1)}+m^2\Psi_Q^{(2)}\cdots, 
\label{eq:expansion}
\end{equation}
\begin{eqnarray}
\Psi_Q^{(0)}&=&\Bigl[\prod_i(1-Q_i)\Bigr]\Psi_{\rm G}, \\
\Psi_Q^{(1)}&=&\Bigl\{\sum_jQ_j\Bigl[\prod_{i(\ne j)}(1-Q_i)\Bigr]\Bigr\} 
\Psi_{\rm G}, \\
\Psi_Q^{(2)}&=&\Bigl\{\sum_{j,k(\ne j)}Q_jQ_k
\Bigl[\prod_{i(\ne j,k)}(1-Q_i)\Bigr]\Bigr\}\Psi_{\rm G}. \qquad 
\end{eqnarray} 
The bases that compose $\Psi_Q^{(\ell)}$ possess necessarily 
$\ell$ lattice sites which are occupied by doublons (or holons) 
without a holon (doublon) in their nearest-neighbor sites. 
Note that in the vicinity of a Mott transition, $g$ 
(Gutzwiller factor) is so small that the doublon density is 
very low ($D\lsim 0.04$, as shown in Fig.~\ref{fig:doublon}). 
Using eq.~(\ref{eq:expansion}), the kinetic and interaction energies 
are formally expanded with respect to $m$, as 
\begin{eqnarray} 
E_{\rm kin}(g,m)&=&K_0+mK_1+m^2K_2+\cdots,\qquad \\
E_{\rm int}(g,m)&=&I_0+m^2I_2+\cdots, 
\end{eqnarray} 
in which $K_\ell$ and $I_\ell$ ($\ell=0,1,2,\cdots$) are functions 
of the Gutzwiller factor $g$ 
\begin{eqnarray}
&&K_0+I_0=E_0=\langle 0|{\cal H}|0\rangle/N_0, \\
&&K_1=\left[\langle 0|{\cal H}_{\rm kin}|1\rangle+
     \langle 1|{\cal H}_{\rm kin}|0\rangle\right]/N_0, \qquad\qquad
\label{eq:K1} 
\end{eqnarray}
$$
K_2+I_2=\Bigl[\langle 0|{\cal H}_{\rm kin}|2\rangle+
     \langle 1|{\cal H}_{\rm kin}|1\rangle+
     \langle 2|{\cal H}_{\rm kin}|0\rangle+\quad
$$
\begin{equation}
     \langle 1|{\cal H}_{\rm int}|1\rangle+
  N_1\langle 0|{\cal H}|0\rangle\Bigr]/N_0,
\end{equation}
with $|\Psi_Q^{(\ell)}\rangle\equiv|\ell\rangle$ and 
$N_\ell=\langle\ell|\ell\rangle$. 
Thus, the total energy is given as 
\begin{equation} 
E(g,m)=E_0+mK_1+m^2\left(K_2+I_2\right)+\cdots. 
\label{eq:total}
\end{equation}
From eq.~(\ref{eq:total}), it seems  that $E(g,m)$ has a minimum 
at a finite $m$ (or $\mu<1$), if $K_1$ is negative. 
Actually, Yokoyama and Shiba\cite{YSMott} calculated $K_1$ for 
one-dimensional systems, and confirmed that $K_1$ is negative. 
They thus concluded that $m$ is always finite as long as $U/t$ is 
finite, that is, a Mott transition does not arise. 
\par

\begin{figure}
\begin{center}
\includegraphics[width=7.0cm,height=2.0cm]{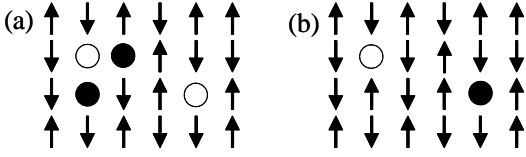}
\end{center}
\vspace{-0.0cm}
\caption{
Examples of electronic configurations appearing (a) in $\Psi_Q^{(1)}$ 
and (b) in $\Psi_Q^{(2)}$, at half filling. 
Solid (open) circles indicate doublons (holons). 
Each example belongs to the case in which the doublon number is 
minimum. 
}
\label{fig:d-h}
\end{figure}

Later, however, this conclusion turned out to be substantially 
incorrect through a more careful study.\cite{YokoPTP} 
Because, at half filling, a doublon and a holon are created 
simultaneously as a pair, a basis included in $\Psi_Q^{(1)}$ has 
at least two doublons, whereas $\Psi_Q^{(2)}$ contains bases with 
only one doublon, as shown in Fig.~\ref{fig:d-h}. 
Thus, as $g$ decreases, $|K_1|$ can become negligible, 
compared with $|K_2|$. 
In fact, $K_1/t$ has been calculated using VMC with 
${\cal P}_Q{\cal P}_{\rm G}\Phi_{\rm F}$ for the one-dimensional 
Hubbard model, with the result that as $g$ decreases, $|K_1|/t$ is 
abruptly reduced from $\sim 0.1$ ($g=0.4$) to $\sim 0.007$ 
($g=0.2$).\cite{YSMott} 
In addition, the quadratic behavior of $E(m,g)$ for small $g$ and 
$m\rightarrow 0$, namely $\partial E/\partial m=0$, is clearly observed 
for two-dimensional cases.\cite{YokoPTP} 
Eventually, it was confirmed that Mott transitions actually take 
place for the square lattice Hubbard model by applying ${\cal P}_Q$ 
to $\Phi_{\rm F}$ \cite{YokoPTP} and $\Phi_d$.\cite{YTOT} 
Afterward, Mott transitions have been observed using ${\cal P}_Q$ 
for various systems in addition to the present model: 
the lattice shown in Fig.\ref{fig:model}(b) with both $\Phi_{\rm F}$ 
and $\Phi_d$,\cite{YTOnew} the kagom\'e  lattice with $\Phi_{\rm F}$, 
\cite{kagome} the checker-board lattice with $\Phi_{\rm F}$, \cite{CB} and 
a degenerate Hubbard model on the square lattice with $\Phi_{\rm F}$. 
\cite{twoband}
\par 

%*****************************************************************************
%  Configurations of \Psi1 and \Psi2
%*****************************************************************************
%*****************************************************************************

Independently of ref.~\citen{YSMott}, Millis and Coppersmith 
\cite{Millis} 
studied the possibility of a Mott transition with a virtually equivalent 
variational wave function to ${\cal P}_Q{\cal P}_{\rm G}\Phi_{\rm F}$. 
To determine whether or not a Mott transition arises, they adopted 
a general theory of Kohn \cite{Kohn} in estimating a zero-frequency part 
of the optical conductivity from the linear response of twisted-boundary 
systems to a small flux. 
They concluded that a wave function like 
${\cal P}_Q{\cal P}_{\rm G}\Phi_{\rm F}$ never becomes insulating 
for a finite $U/t$, on the basis of a VMC result with a 
nonoptimized $g(=0.3)$ for a one-dimensional small (10-site) system. 
However, this conclusion is an error, as we have demonstrated with actual 
counterevidence in the preceding paragraph (or in \S\ref{sec:mott}). 
We need to check this study as to which process causes the error. 
\par

%-------
\subsection{Comments on recent VMC study\label{sec:Liu1}}
%-------

%*****************************************************************************
% results of J. Liu
%*****************************************************************************
\begin{figure}
\begin{center}
\includegraphics[width=8.5cm,height=10.0cm]{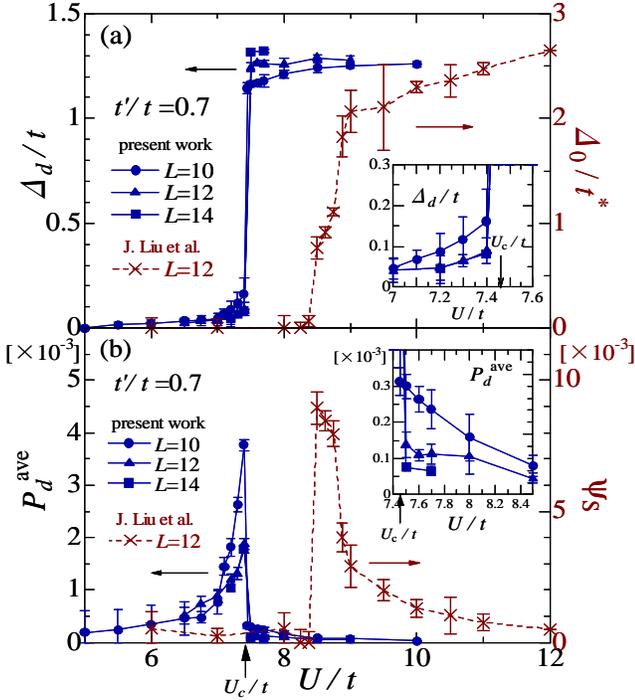}
\end{center}
\vspace{-0.5cm}
\caption{(Color online) 
Comparison of VMC results for $t'/t=0.7$ between the study of 
Liu \etal\ (taken from Fig.~1 in ref.~\citen{Liu}) and the present study 
for (a) the $d$-wave gap (the optimized variational parameters), 
and (b) the pairing correlation function. 
The quantities studied by Liu \etal, $\Delta_0/t^*$ and $\psi_{\rm s}$, 
(crosses) practically correspond to $\Delta_d/t$ and $P_d^{\rm ave}$ 
in the present study (solid symbols), respectively. 
}
\label{fig:liu}
\end{figure}
%*****************************************************************************

Recently, Liu \etal\ \cite{Liu} have studied the same model eq.~(\ref{eq:model}) 
using a VMC method with an almost identical $d$-wave singlet 
trial function (abbreviated as $\Psi_{\rm Liu}$) to ours, $\Psi_Q^d$, 
in eq.~(\ref{eq:wf}). 
The differences between $\Psi_{\rm Liu}$ and $\Psi_Q^d$ are as follows: 
(i) In $\Psi_{\rm Liu}$, to improve the renormalization effect of the 
quasi-Fermi surface, $\tilde t''$ is additionally introduced in the 
other diagonal $(-1,1)$ direction in $\tilde\varepsilon_{\bf k}$ 
as a parameter. 
(ii) In $\Psi_{\rm Liu}$, the magnitude of the doublon-holon binding 
factors for next-nearest-neighbor sites is assumed identical to 
that for the nearest-neighbor ones, namely $\mu'=\mu$ in our notation. 
Because these differences are not essential, \cite{notediff} it is natural 
that the results of Liu \etal\ are not profoundly different from 
the present ones, and that some of their arguments are common to ours; 
we are ready to respect this aspect of their study. 
However, in a crucial point in connection with a Mott transition, 
their interpretation is distinct from ours and questionable. 
Because it is vital to clarify the properties of Mott transitions 
in this type of wave function for related studies, it is necessary 
to clarify the questionable points. 
\par

In the study of Liu \etal, a pair correlation function ($\psi_s$) was 
calculated for $t'/t=0.7$ with a fixed system size, $L=12$. 
We have traced $\psi_s$ from Fig.~1 in ref.~\citen{Liu}, and plotted 
it in Fig.~\ref{fig:liu}(b) together with our corresponding results. 
For small values of $U/t$, $\psi_s$ is negligible, whereas it 
increases at approximately $U/t=8.5$ and then it gradually decreases. 
This behavior of $\psi_s$ is interpreted in ref.~\citen{Liu} as follows: 
For small values of $U/t$, the state is a conventional metal; 
robust SC occurs at approximately $U/t=8.5$, and {\it weak SC continues 
to larger values of $U/t$ without a transition}. 
The problematic interpretation consists in the last point. 
In comparing $\psi_{\rm s}$ with $P_d^{\rm ave}$, 
$U/t$ at which $\psi_s$ becomes conspicuously large ($\sim 8.5$) 
is larger roughly by 1 than $U_{\rm c}/t$ (=7.45) of $P_d^{\rm ave}$. 
This discrepancy may stem from point (ii) of the differences 
in the wave functions, \cite{notediff} and is of little importance. 
The most crucial point is that they overlooked the Mott transition 
this type of wave functions exhibits immediately above the region 
of robust SC. 
They probably accepted the incorrect claim of Millis and Coppersmith 
mentioned in \S\ref{sec:exp}, and neglected to check the system-size 
dependence of $\psi_{\rm s}$. \cite{noteLdep}
For such a finite system as $L=12$, $\psi_{\rm s}$ is still visibly 
finite, but must decrease abruptly with increasing $L$, as we showed 
the system-size dependence of $P_d^{\rm ave}$ in the inset of 
Fig.~\ref{fig:liu}(b). 
\par

This misinterpretation influences some other points. 
Because the region of SC is considered to continue to larger 
values of $U/t$ ($\gsim 8.7$), the features of the insulating 
state are confused with those of the SC state. 
Although the prominent renormalization of $\varepsilon_{\bf k}$ 
(or $t'$ and $t''$) is considered to occurs in the SC state, 
as shown in Fig.~3 of ref.~\citen{Liu}; correctly, the renormalization 
of $t'/t$ is small in the SC state, but prominent in the insulating 
state, as we showed in \S\ref{sec:mott} and \S\ref{sec:pair}. 
Furthermore, the spin correlation function is compared between 
the metallic regime ($U/t=7$), for which $\psi_{\rm s}$ is negligibly 
small, and the insulating regime ($U/t=11$) in Fig.~4 of ref.~\citen{Liu}. 
Therefore, as long as the spin correlation in the {\it SC state} is 
concerned, this comparison is useless. 
\par

Finally, we point out the possible fictitious enhancement of 
$P_d^{\rm ave}$ (or $\psi_{\rm s}$) specific to finite system sizes. 
In Fig.~\ref{fig:liu}(b), the values of $P_d^{\rm ave}$ immediately 
below $U_{\rm c}$ for $L=10$ are roughly twice larger than those 
for $L=12$ and 14. 
This is because the $k$-point configuration for $L=10$ happens to be 
specially advantageous for SC under the given condition, including 
the renormalized effect of $\varepsilon_{\bf k}$ by $U$. 
This is confirmed by the fact that the quasi-Fermi surface has a 
different shape only in these special cases (not shown). 
Undoubtedly, the behavior of $P_d^{\rm ave}$ for $L=12$ and 14 will be
closer to that for $L=\infty$. 
In accordance with this anomalous enhancement in $P_d^{\rm ave}$, 
$\Delta_d/t$ takes large values as observed in the inset of 
Fig.~\ref{fig:liu}(a). 
According to our experience, \cite{YTOnew} such behavior is not rare 
for the marginal values of $t'/t$ where SC is vanishing. 
When we return to the data of $\Psi_{\rm Liu}$ in Fig.~\ref{fig:liu}(a) 
with this situation in mind, we notice that the values of $\Delta_0/t^*$ 
for $U/t\sim 8.5$ (three points), where $\psi_{\rm s}$ is considerably 
enhanced, are specifically increased. 
Hence, the prominent peak in $\psi_{\rm s}$ at $U/t\sim 8.5$ is probably 
a fictitious increase specific to their system size and boundary 
conditions. 
\par

We hope improved studies of $\Psi_{\rm Liu}$ will appear. 
\par

%%%%%%%%%%%%%%%%%%%%%%%%%%%%%%%%%%%%%%%%%%%%%%%%%%%%%%%%%%%%%%%%%%%%%%%%%

%%%%%%%%%%%%%%%%%%%%%%%%%%%%%%%%%%%%%%%%%%%%%%%%%%%%%%%%%%%%%%%%%%%%%%%%%


\begin{thebibliography}{99}
\bibitem{comment1} At half filling, the results for $t'(<0)$ are 
identical to those for $|t'|$. 
We hence concentrate on the case of $t'/t\ge 0$.
\bibitem{commentPQ} Because the $t$-$J$ and Hubbard models are related 
in the strong coupling regime through a canonical transformation, 
${\cal H}_{t-J}\approx e^{iS}{\cal H}_{\rm Hub}e^{-iS}$, \cite{Harris}
in which $S$ is given as
$$
\sum_{\langle ij\rangle\sigma}\frac{-it}{U}
\left[n_{i-\sigma}(1-n_{j-\sigma})c^\dagger_{i\sigma}c_{j\sigma}
     -n_{j-\sigma}(1-n_{i-\sigma})c^\dagger_{i\sigma}c_{j\sigma}\right], 
$$
the expectation values of the $t$-$J$ model with respect to $\Psi_{\rm G}$ 
are approximately equivalent to those of the Hubbard model with respect to 
$e^{-iS}\Psi_{\rm G}$. 
When one expands $e^{-iS}$, the first-order part ($\sim t/U$) of 
$e^{-iS}\Psi_{\rm G}$ broadly corresponds to ${\cal P}_{\rm Q}\Psi_{\rm G}$. 
Thus, the favorable feature in treating the $t$-$J$ model with 
$\Psi_{\rm G}$ is shared with the case in which the Hubbard model is 
treated with ${\cal P}_{\rm Q}\Psi_{\rm G}$. 
\par
\bibitem{noteZ} It is difficult to estimate proper values of $Z$ 
as a function of $U/t$ with a finite-sized system, in the case that 
the position of the quasi-Fermi point moves and crosses a $k$ point, 
when $U/t$ is varied. 
As an example, we show the case of $t'/t=0.4$ and $L=12$ in 
Fig.~\ref{fig:ren}, where the data for $U/t=6.5$ deviates to some extent. 
For this parameter set, we find the quasi-Fermi point to shift in the 
nodal direction in Fig.~\ref{fig:mom}(a).
\bibitem{notesize} Generally, as system size decreases, 
the critical behavior near a phase transition tends to be less sharp, 
because a transition, in a strict sense, takes place only for
$N_{\rm s}\rightarrow\infty$. 
For instance, the curve for $t'=0$ and $L=10$ in Fig.~\ref{fig:doublon} 
(open square) is relatively smooth, whereas the curve for $L=16$  
seems to have an obvious jump. 
Thus, one must necessarily check system-size dependence to treat 
a phase transition.
\bibitem{noteregime} The weak correlation regimes, we here assume, 
are $U/t<6$ for $t'/t=0.2$, 0.4, and 0.6, and $U<U_{\rm c}$ for 
$t'/t=0.7$ and 0.8.
\bibitem{notediff} The effect of difference (i), namely of 
$\tilde t''/t$, must be slight, because, according to Fig.~2 in 
ref.~\citen{Liu}, the optimized $|\tilde t''/t|$ is small 
both for the conductive ($U/t\lsim 8.5$) and insulating ($U/t\gsim 8.5$) 
phases. 
The effect of difference (ii), namely of putting $\mu'=\mu$ in 
$\Psi_{\rm Liu}$, manifests itself to some extent, because, as in 
Figs.~\ref{fig:para}(d) and \ref{fig:para}(e), the optimized values 
of $\mu$ and $\mu'$ are quite different, especially in the insulating 
phase. 
Thus, it is possible that $\Psi_{\rm Liu}$ becomes quantitatively 
worse than $\Psi_Q^d$ for $U\gsim U_{\rm c}$; better results would 
be obtained, if $\mu'=0$ is used instead of $\mu'=\mu$ in $\Psi_{\rm Liu}$.
\bibitem{noteLdep} In ref.~\citen{Liu}, the system-size dependence 
of the gap parameter $\Delta_0/t^*$ for $U/t=9.5$ (in the insulating 
regime) is shown in the inset of Fig.~1. 
Because the finite $\Delta_0/t^*$ does not necessarily indicate 
the realization of SC, as we have mentioned, it is useless to check 
whether $\Delta_0/t^*$ remains finite for $L\rightarrow\infty$, 
in order to determine whether or not SC arises.
\end{thebibliography}
\end{document}